\documentclass[11pt,a4paper]{article}

\usepackage[mathscr]{euscript}
\usepackage[textsize=scriptsize,textwidth=2.5cm]{todonotes}
\usepackage{amssymb,amsmath,latexsym,bm,amsfonts,mathtools}
\usepackage{graphicx}
\usepackage{longtable}
\usepackage{color}
\usepackage{indentfirst}
\usepackage{subfig}
\usepackage{comment}
\usepackage[normalem]{ulem}
\usepackage{array}
\usepackage{tikz-cd}
\tikzcdset{scale cd/.style={every label/.append style={scale=#1},
    cells={nodes={scale=#1}}}}

\usepackage{caption} 
\usepackage{floatrow}

\usepackage{multirow}
\usepackage{booktabs}
\usetikzlibrary{positioning, shapes.geometric, patterns, graphs, decorations.pathmorphing}
\usepackage{relsize}
\usepackage{graphicx}
\usepackage{calc}

\numberwithin{equation}{section}

\newcommand{\Z}{\ensuremath{\mathbb{Z}}}

\newcommand{\bO}{\boldsymbol{O}}

\newcommand{\bxi}{\boldsymbol{\xi}}
\newcommand{\bpsi}{\boldsymbol{\psi}}

\newcommand{\eq}[1]{\begin{align}#1\end{align}}

\def \p {\partial}
\def \ba {\begin{array}}
\def \ea {\end{array}}
\def \defa {\beta}
\def \defb {\gamma}

\newcommand{\tof}{A}

\newcommand{\tofp}{q}
\newcommand{\vvfp}{\nu}

\begin{document}

\title{A BMS-invariant free fermion model}
\author{
Peng-xiang Hao$^{a}$,
Wei Song$^{a,b}$,
Zehua Xiao$^a$,
and Xianjin Xie$^a$\footnote{pxhao@mail.tsinghua.edu.cn,  wsong2014@mail.tsinghua.edu.cn,\\ xiaozh20@mails.tsinghua.edu.cn, xxj19@mails.tsinghua.edu.cn}
}

\date{}
\maketitle
\begin{center}
{\it
$^{a}$Yau Mathematical Sciences Center, Tsinghua University, Beijing 100084, China\\

$^{b}$Peng Huanwu Center for Fundamental Theory, Hefei, Anhui 230026, China}
\vspace{10mm}
\end{center}

\begin{abstract}

We used the Cartan formalism to construct fermionic models that are compatible with Galilean or Carrollian symmetry and rigid scaling symmetry. The free Carrollian fermion model exhibits conformal Carrollian symmetry which is isomorphic to  the asymptotic symmetries for flat spacetime in three dimensions, namely the BMS$_3$ symmetry. We performed canonical quantization to this free BMS fermion model, discussed both the highest weight vacuum and the induced vacuum, calculated the correlation functions and the torus partition function. Finally we constructed $\mathcal N=2$ supersymmetric theories by combining the free fermion model and the free scalar model \cite{Hao:2021urq}

\end{abstract}

\baselineskip 18pt
\thispagestyle{empty}

\newpage
\tableofcontents
\newpage

\section{Introduction}
Symmetry, if exists, always plays important roles in modern theoretical physics.
Interestingly, the study of the symmetry of Einstein gravity on asymptotically flat spacetime \cite{Bondi:1960jsa, Bondi:1962px, Sachs:1962wk, Sachs:1962zza} lead to  two surprises: one is that the resulting asymptotic symmetry group, the BMS group, is infinite dimensional and hence is much larger than  the isometry group of Minkowski spacetime, namely the Poincar\'{e} group; the other is that for a long time the BMS symmetry had seemed very mysterious and it had not been clear what useful lessons one can draw from the symmetry, until some recent progresses. Recently, the BMS symmetry has been related to Weinberg's soft theorem \cite{Strominger:2013lka} and gravitational memory effect \cite{Strominger:2014pwa, Strominger:2017zoo}. Relatedly scattering amplitude in flat spacetime is related to a two dimensional CFT called celestial CFT \cite{Pasterski:2016qvg,Pasterski:2017kqt, Pasterski:2021rjz,Pasterski:2021raf,Donnay:2020guq,Prema:2021sjp}, and moreover connections between the BMS group and celestial CFTs have been discussed in \cite{Donnay:2022aba,Bagchi:2022emh}.

Parallel to the developments related to BMS group in four dimensions, recently the study of BMS group in three dimensions has also been fruitful.  The story is closely related to the bottom-up approach of the AdS$_3$/CFT$_2$ correspondence. For three dimensional gravity with a negative cosmological constant, the phase space of gravity can be organized into representations of its asymptotic symmetry group under some consistent boundary conditions, which turns out to be the two dimensional conformal group \cite{Brown:1986nw}. The rich results of two dimensional conformal field theories (CFT$_2$) enable many questions to be addressed in a more explicit and precise way,  including the microscopic counting of the black hole entropy, correlation functions, quantum entanglement, etc.
Using a similar strategy to three dimensional gravity without cosmological constant, the asymptotic group of flat spacetime is the three dimensional BMS group (BMS$_3$), which is also infinite dimensional \cite{Ashtekar:1996cd,Barnich:2006av,Barnich:2012aw}.
The BMS$_3$ algebra is found to be isomorphic to two dimensional  Carrollian conformal algebra, which can be obtained from left and right moving Virasoro algebras by taking an ultra-relativistic limit \cite{Lvyleblond1965UneNL}, hence the name.
In analogy to the AdS$_3$/CFT$_2$ correspondence, one can try to set up a holographic equivalence between asymptotically flat spacetime and BMS$_3$-invariant field theory (BMSFT), or equivalently Carrollian conformal field theory (CCFT)\cite{Duval:2014uva,Duval:2014uoa,Duval:2014lpa}.
Note that it happens that the Carrollian conformal algebra in two dimensions is also isomorphic to the Galilean conformal algebra, namely the non-relativisitic limit of the Virasoro algebra \cite{contraction}. Due to this, Galilean conformal field theory (GCFT) was often used in the early literature of flat holography \cite{Bagchi:2009my,Bagchi:2010zz}.  Although GCFT and CCFT share a lot of similar features, the differences are also important, especially on higher dimensions \cite{Chen:2021xkw, Figueroa-OFarrill:2022nui}.
Many interesting results have been obtained along this direction of flat holography. Using the symmetry, correlation functions for BMSFT and GCFT have been analyzed \cite{Bagchi:2009ca,Saha:2022gjw,Chen:2020vvn}, and BMS bootstrap program was initiated in \cite{Bagchi:2016geg,Bagchi:2017cpu,Chen:2020vvn,Chen:2022cpx,Chen:2022jhx}. Geodesic Witten diagrams were considered in \cite{Hijano:2017eii}.
The phase space of Einstein gravity contains the so-called flat cosmological solutions featuring Cauchy horizons on which an entropy can be assigned \cite{Barnich:2012xq}.
The torus partition function of BMSFTs was found to be modular invariant and a Cardy-like formula can be derived, so that it provides an interpretation of the entropy of the Cauchy horizon \cite{Bagchi:2012xr}.
As an important measure of quantum entanglement, entanglement entropy for the BMSFTs were calculated \cite{Bagchi:2014iea,Basu:2015evh}, and the holographic dual was proposed \cite{Jiang:2017ecm, Apolo:2020bld}. Other interesting topics including modular Hamiltonian \cite{Apolo:2020qjm},  entanglement negativity, the reflected entropy \cite{Basu:2021awn}, and partial differential entropy \cite{Camargo:2022mme} have also been discussed in the literature.

Despite these developments, not much is known about the putative dual BMSFTs other than the properties implied by the symmetries. Therefore it is necessary to construct explicit models of quantum field theories with BMS symmetries.
So far a Liouville-like theory with BMS symmetries has been constructed and discussed in \cite{Barnich:2012rz,Barnich:2013yka,Bagchi:2021gai}, which can be obtained from the ultra-relativistic (UR) limit of the ordinary Liouville theory, or from the geometric action of the BMS$_3$ group \cite{Barnich:2017jgw}; a free scalar BMS model was analyzed in \cite{Hao:2021urq}, whose action also comes from the tensionless limit of string theory \cite{PhysRevD.16.1722,Isberg:1993av}, and a $\sqrt{T\bar{T}}$ deformation of a free scalar CFT$_2$ \cite{Rodriguez:2021tcz,Bagchi:2022nvj,Tempo:2022ndz}; and more general models with Galilean or Carrollian symmetry in two dimensions and higher dimensions  have also been proposed in \cite{Bagchi:2019xfx,Bagchi:2019clu,deBoer:2021jej,Bagchi:2022eav,Rivera-Betancour:2022lkc,Baiguera:2022lsw,Banerjee:2022uqj}.

In this paper, we study two dimensional free fermion models with BMS symmetry, or equivalently Carrollian conformal symmetry.
In two dimensions, both the Carrollian and Galilean transformations consist of two translational and a boost transformation,
\eq{
x^a&\rightarrow x^a+\delta^a,\, \quad a=1,2\\
x^2&\rightarrow x^2+v x^1}
where the time direction is chosen to be $x^1$ for Galilean symmetry, and $x^2$ for Carrollian symmetry, and $v$ is a boost parameter.
By keeping the choice of time implicitly, we can discuss both the symmetries in a uniform way.
Using the Cartan formalism, we systematically search for possible fermionic theories invariant under Galilean or Carrollian symmetry together with rigid scaling in two spacetime dimensions.
In order for a spinor with two Majarona components to couple consistently with a flat Galilean or Carrollian geometry, there are two possibilities: i) the two fermions are decoupled free chiral fermions, or ii) the kinetic term involves products of the the two Majarona fermions and hence they are not decoupled. The latter is the main focus of this paper.  After further specifying the Carrollian case, we find that the model exhibits BMS symmetry, and thus is referred to as the BMS fermion model.
The action on the cylinder is given by
\begin{align}
S =- \frac{i}{2\pi}\int d\tau d\sigma (\psi_1\partial_{\sigma}\psi_1-\psi_2\partial_{\tau}\psi_1),\label{BMSf}
\end{align}

which agrees with the action of the worldsheet fermion in the tensionless limit with inhomogeneous scaling \cite{Bagchi:2017cte}. 

We then perform canonical quantization,  and compute the correlation functions and torus partition functions in both the highest representation and the induced representation.
In the highest weight representation, the BMS algebra has central charges $c_L=1,\, c_M=0$. The total torus partition which includes contributions from all the R-R, NS-NS, R-NS and NS-R sectors is invariant under the modular $S$ transformation, but picks up a phase under the modular $T$ transformation. It is possible to restore  modular $T$ invariance if we take 24 copies of this BMS fermion theory.
In the induced representation, the central charges of the BMS algebra both vanish, the correlation function is ultra-local, and the partition functions is divergent.

Finally, we construct supersymmetric theories with BMS symmetry.
Combining the BMS scalar discussed in \cite{Hao:2021urq} and the aforementioned BMS fermion together, we get a $\mathcal N=2$ supersymmetric algebra with BMS algebra as the bosonic part.
Alternatively, combining the BMS scalar and with two chiral fermions also lead to  $\mathcal N=2$ supersymmetry, which contain two  $\mathcal N=1$ subalgebra.  This means that we can also build an $\mathcal N=1$ theory using the BMS scalar together with one chiral fermion.

The layout of this paper is as follows. In section 2, we review the Galilean/Carrollian geometry and construct covariant fermionic actions on it. In section 3, we solve the BMS fermion by canonical quantization in the highest weight vacuum. In section 4, we organize the states into the BMS staggered module. In section 5, we calculate the torus partition function in the highest weight representation. In section 6, we consider another choice of the vacuum, the so-called induced vacuum. In section 7, we put the BMS scalar and the BMS fermion together to discuss the BMS supersymmetry.

{\bf Note added: } During the preparation of this draft, we became aware of the paper \cite{Yu:2022bcp} which has overlaps with our results in section 3, section 4 and section 5.

\section{Fermions on the Galilean/Carollian geometry}

In this section we construct free Fermionic models with Galilean or Carrollian symmetry, together with rigid scaling symmetry in the covariant formalism. The discussions in this section apply to both the Galilean and the Carrollian symmetry, which we will refer to as G/C for brevity.
We will review the Cartan formalism with local G/C symmetry, write down the Clifford algebra, and finally use bi-spinors to construct fermionic models with G/C symmetry.

\subsection{Flat Newton-Carrollian geometry}
In this subsection, we briefly review the Cartan formalism with G/C symmetry, following the conventions of \cite{Hofman:2014loa}.  The formalism was first introduced to study non-relativistic theories, and hence is often referred to as the Newton-Cartan geometry in the literature. However, it can similarly be formulated for ultra-relativistic theories as well.
We start from flat geometry compatible with G/C symmetry, and then construct curved geometry by making the flat geometry the tangent space at each point.

Let us choose the two dimensional coordinates as $x^a,a=1,2$.
We are interested in a flat geometry with translational, boost, and scaling symmetry
\eq{
x^a&\rightarrow x^a+\delta^a,\, \quad a=1,2 \nonumber\\
x^a&\rightarrow \Lambda^a\,_b x^b, \label{isometry}\\
x^a&\rightarrow \lambda x^a ,  \nonumber
}
where $\delta^a$ denotes translation along the direction of $x^a$, and $\Lambda$ can be chosen as
\begin{equation}\Lambda=
\begin{pmatrix}
  1 & 0 \\
  v & 1
\end{pmatrix}\label{boost}
\end{equation}
where $v$ is a boost parameter.
In relativistic theory, a boost transformation mixes the time and spatial direction. In contrast, here the boost transformation \eqref{boost} leaves
the $x^1$ direction invariant, and only changes the $x^2$ direction.
If we choose $x^1$ as the time direction, $\Lambda$ is the non-relativistic boost with $v$ being the relative velocity between the two frames, which can be obtained from the Lorentzian boost by sending the speed of light to infinity. In this case, $x^1$ is the ``absolute'' temporal direction. Alternatively, if $x^2$ is chosen as the time direction,  $\Lambda$ corresponds to ultra-relativistic boost which can be obtained by sending the speed of light to zero.\footnote{The warped CFT (WCFT) discussed in \cite{Hofman:2014loa} has the same translation and boost symmetry as in \eqref{isometry}, but with an anisotropic scaling $x^1\to \lambda x^1, x^2\to x^2$.  WCFTs have ``absolute spatial direction" as $x^2$ is usually chosen as time. } In this case, $x^1$ direction is the ``absolute'' spatial direction.  In the following,  most of the discussions are independent of the choice of time direction, and hence we will discuss both limits collectively.
To construct field theories compatible with the symmetry \eqref{isometry}, we need to find invariant tensor and spinor representations of the boost symmetry.

Let us start with tensors with rank one.
Similar to usual differential geometry, we label vectors with upper indices, and one-forms with lower indices.
The boost invariant vector and one-form are defined to satisfy
\begin{align}
\Lambda^a_{\ b}\tofp^b=\tofp^a,\quad
\tofp_b\Lambda^b_{\ a}=\tofp_a
\end{align} We will see later that they are dual to each other, the reason why we use the same notation $\tofp$.
With the choice \eqref{boost}, the explicit solution to the above eigenvalue problem is \begin{equation}
\tofp^a=\left(
\begin{aligned}
& 0\\
& 1
\end{aligned}
\right )\ ,\ \ \ \ \tofp_a=(1\ \ \ 0),\ \ \ a=1,2.\label{qa}
\end{equation}

At rank two, the boost invariant Matrix with two lower indices should satisfy
\begin{equation}
M_{ab}=\Lambda^{a'}\,_{a} M_{a'b'}\Lambda^{b'}\,_b.\label{boostinv}
\end{equation}
Using the 1-form $q_a$, it is straightforward to construct a boost-invariant and symmetric 2-tensor $g_{ab}=q_aq_b$.
Any boost invariant 2-tensor with lower indices can thus be decomposed into a symmetric part $g_{ab}$ and an anti-symmetric part $\epsilon_{ab}$,
\eq{\label{metric}
g_{ab}
=\tofp_a\tofp_b,\ \quad
\epsilon_{ab}=\left(
\begin{aligned}
0\ \ & 1\\
-1\ \ & 0
\end{aligned}
\right )
}
Similar to flat Minkowski spacetime, $g$ is interpreted as a flat metric with Galilean or Carrollian symmetry, which can be used to define inner products between two arbitrary vectors  $U^a$ and $ V^b$,
\eq{ U\cdot  V=U^a g_{ab} V^b\label{innerUV}}
Unlike the Minkowski metric, however,
the non-Minkowski flat metric \eqref{metric} is non-invertible, and hence can not be used to raise and lower indices.
Nevertheless,  we can use the invertible antisymmetric tensor $\epsilon_{ab}$ to map a vector to a one-form. In particular, the boost-invariant one-form $q_a$ is indeed dual to the boost-invariant vector $q^a$,
\eq{\label{lower}
\tofp_a=\epsilon_{ab}\tofp^b, \quad
 \tofp^a=\epsilon^{ab}\tofp_b,\quad \epsilon^{ab} \epsilon_{bc}=\delta^a_{\,c}.
}
Using $\epsilon^{ab}$, we can also define a metric with upper indices from that with lower indices,
\begin{equation}
g^{ab}\equiv\tofp^a\tofp^b=\epsilon^{ac}\epsilon^{bd}\tofp_c \tofp_d=\epsilon^{ac}\epsilon^{bd}g_{cd}.
\end{equation}
Note that the inner product between two vectors can also be written as that of two one-forms, but is not equal to the contraction between a vector and a one-form,
\eq{
U\cdot  V=U^a g_{ab} V^b=U_a g^{ab} V_b\neq U^a V_a=-U_a V^a\,.
}
where the minus sign in the last equality is due to the anti-symmetric property of $\epsilon$. In particular, the contraction of a vector with its dual 1-form always vanishes, namely  \eq{V^a V_a=0.\quad }

Note that the boost symmetry has selected a direction, specified by the boost invariant one-form $q_a$, or equivalently, the boost invariant vector $q^a$.
Using the explicit expression \eqref{qa}, this corresponds to selecting the ``absolute'' direction $x^1$.
Most of our discussions do not depend on the explicit expression \eqref{qa}, but only on its properties.
The vector $q^a$ is orthogonal to an arbitrary vector in the two dimensional vector space under the inner product \eqref{innerUV}, namely
\eq{q\cdot V =q^a g_{ab} V^b=0\label{qV}}
In the two dimensional vector space, we can choose a basis $(q^a, \, \nu^a)$ by requiring $\nu^a$ to be linearly independent of $q^a$, or equivalently
 \eq{
 N=\vvfp^a\tofp_a\neq 0. }

So far we have only considered the boost symmetry. Now let us turn to the scaling symmetry \eqref{isometry}.
Define the scaling structure $J^a_{\ b}$ as the generator of the infinitesimal scaling,
so that the finite transformation $e^{-\lambda J} $ acts as \eqref{isometry}.
Then the action of $J^a_{\ b}$ on the basis $(q^a, \, \nu^a)$ of the vector space has to be \begin{equation}
J^a_{\ b}\tofp^b=-\tofp^a,\ \ \ J^a_{\ b}\vvfp^b=-\vvfp^a
\end{equation}
Using the basis $\tofp,\vvfp$ and the anti-symmetric tensor $\epsilon$, the scaling structure can be expressed in a coordinate-independent way as,
\begin{equation}
J^a_{\ b}=-N^{-1}(\vvfp^a\tofp_b-\tofp^a\vvfp_b)\label{scaling}
\end{equation}

To summarize this subsection, flat geometry compatible with translational invariance and boost invariance can be specified by
 a boost invariant one-form $q_a$, and a boost invariant antisymmetric tensor $\epsilon^{ab}$.
The degenerate flat metric $g_{ab}=q_aq_b$ defines the inner product between vectors \eqref{innerUV}, while the antisymmetric tensor $\epsilon^{ab}$ and its inverse $\epsilon_{ab}$ raises and lowers indices. This is in contrast with  the usual Riemannian geometry, in which case the metric plays both roles. The boost invariant 1-form $q_a$ selects an ``absolute'' direction.
Further considering an isotropic scaling symmetry \eqref{isometry}, we can express the scaling structure as \eqref{scaling} which is coordinate independent.

\subsection{Connection on Galilean/Carrollian geometry}
In this subsection we will discuss curved Galilean/Carrollian geometry in Cartan formalism. In particular, we will find the affine connection and spin connections compatible with local boost symmetry and rigid scaling symmetry \eqref{isometry}.

In the Cartan formalism,  the tangent space at each point of a curved manifold is the aforementioned flat geometry with Galilean/Carrollian symmetry. Thus the 2d Newton/Carrollian geometry can be described by the data $(M,\tofp_a,e^a\,_\mu,\epsilon_{ab})$, where $M$ is a 2-dimensional manifold, $\tofp_a$ is the boost invariant one-form in the cotangent space, $\epsilon_{ab}$ is the anti-symmetric tensor that maps one-forms to the vectors, and the vielbein $e^a\,_\mu$ maps space-time vectors to tangent space vectors,
\begin{equation}
e^a_{\ \mu}:v^{\mu}\rightarrow {v}^a
\end{equation}
and is assumed to be invertible. Similar to the discussion of Riemannian geometry, covariant derivative is defined as
\begin{equation}
D=\partial+\omega-\Gamma\label{covD}
\end{equation}
where $\omega$ is the spin connection which acts on the tangent space indices $a,b,\cdots$, while $\Gamma$ is the affine connection acting on spacetime indices $\mu,\nu\cdots$.
The torsion and curvature two-forms are respectively
\begin{equation}
T^a=de^a+\omega^a_{~b} \wedge e^b, \ \
R^a_{~b}=d\omega^a_{~b}.
\end{equation}

As discussed at the end of section 2.1, flat geometry is specified by the boost invariant vector $q^a$ which selects an absolute direction in flat geometry, and the antisymmetric 2-tensor $\epsilon_{ab}$ which lowers indices.
 We would like to find a spin connection that keeps $q^a$ and $\epsilon_{ab}$ covariantly constant.
First we note that requiring
\begin{equation}
D_\mu\epsilon_{ab}=0\label{de0}
\end{equation}
implies that the spin connection one form $\omega_{ab}$ is symmetric in the two lower indices, namely
\begin{equation}
\omega_{ab}=\omega_{ba}.
\end{equation}
Further using \eq{D_\mu q^a=0\label{de1}}
we find that
 the spin connection $\omega^a_{~b\mu}$ can be written as,
\begin{equation}
\omega^a_{~b\mu}=\tofp^a\tofp_b\omega_{\mu}.\label{spin}
\end{equation}
Note that \eqref{de0} and \eqref{de1} also imply that the boost-invariant one-form is also covariantly constant, $D_\mu q_a=0.$
One can verify that the torsion free condition does not further specify the spin connection, similar to the observation for the affine connection \cite{Hartong:2015wxa}.
This is in contrast with the relativistic case,
where the affine connection is determined uniquely by the torsion free and the metric compatibility conditions.
Then the spin connection can be determined from the vielbein postulate together with the fact that it is anti-symmetric.
In the non-Lorentzian case, however, it has been observed that the aforementioned procedure cannot determine the spin connection uniquely, and additional conditions are needed \cite{Jensen:2014aia,Bergshoeff:2014uea}.
In the following, we will just keep the general form of the spin connection.

To end this subsection, we comment on an alternative formulation of the G/C geometry. Using the vielbein, we obtain a spacetime one-form $\tof_\mu=\tofp_ae^a_\mu$ and a symmetric 2-tensor $G^{\mu\nu}=g_{cd}\epsilon^{ac}\epsilon^{bd}e_a^\mu e_b^\nu$. Similar to the property \eqref{qV} of $\tofp_a$,  the spacetime vector $A$ is also orthogonal to an arbitrary spacetime one-form, as $\tof_\mu G^{\mu\nu} v_\mu=0$. In the Galilean case, $\tof$ is called the temporal one form or the clock one form, and $G^{\mu\nu}$ is called the inverse spatial metric \cite{Bekaert:2014bwa}. In the Carrollian case, $\tof$ is now the one-form in the spatial direction, and the metric is degenerate in the time direction, so that $G^{\mu\nu}$ is the inverse temporal metric. The geometry can thus be described by the triplet $(M, \,A,\,G)$.
This is the analog of describing Riemann geometry using coordinates.
In the discussion of connection, the two formalisms are compatible.
In order to discuss fermions, however, we need to use the Cartan formalism which allows us to couple the Fermions to curved geometry.

\subsection{Fermions on (flat) Galilean/Carrollian geometry}\label{sec.2.3}
In this subsection we couple fermions to flat Galilean/Carrollian geometry, and construct an action compatible with the local G/C symmetry.
 In order to do so, we need to define
a Clifford algebra which is a spinor representation of the symmetry \eqref{isometry}, a charge conjugation operator which defines the dual spinor space, and $\Gamma^*$, the analog of $\Gamma^5$, which transforms under the local G/C group as a psuedoscalar. With these definitions, we can build bi-spinors which transform invariantly or covariantly under the action of G/C transformation, and further construct fermionic actions from these bi-spinors.

In Galilean/Carrollian geometry, we can define
 the Clifford algebra by requiring the anti-commutation relation
\begin{equation}
\{\Gamma^a,\Gamma^b\}=2g^{ab}=2\tofp^a\tofp^b\label{gammadef}
\end{equation}
which is similar to relativistic theories with the Riemannian metric $g^{ab}$.
Using \eqref{gammadef}, we can define a boost generator $M_0$ which generates the boost \eqref{isometry} on spinors,
\eq{
M_0={1\over8}\epsilon_{ab}[\Gamma^a,\Gamma^b]={1\over4}\epsilon_{ab} \Gamma^a\Gamma^b
}
so that the Gamma matrices transform under the boost transformation as a vector, namely
\eq{
[M_0,\,\Gamma^a]=\tofp_b\,\tofp^a\,\Gamma^b\,.
}
In two spacetime dimensions, the Gamma matrices acts on spinor field $\psi\equiv\psi_\alpha$,
\begin{equation}
\psi=\left(
\begin{aligned}
& \psi_1\\
& \psi_2
\end{aligned}
\right )\label{spinor}
\end{equation}
In order to build boost invariant bi-spinors, we need to first find a dual spinor $\bar{\psi}$ which has the following behaviour under the infinitesimal boost,
\begin{equation}
\overline{(M_0\psi)}=-\bar{\psi}M_0.
\end{equation}
There are two ways to  define $\bar \psi$, either through the charge conjugation matrix $C$ or the Dirac conjugation matrix $D$, defined respectively by
\begin{equation}\label{conjugation}
C\Gamma^a=\pm(\Gamma^a)^TC,\ \ \ D\Gamma^a=\pm(\Gamma^a)^\dagger D.
\end{equation}
Given a solution of \eqref{gammadef}, one can solve the above equations
to find the charge conjugation matrix $C$, and the Dirac conjugation matrix $D$. As we will show later, only the minus sign of the above equation allows non-trivial solutions. This is in contract with relativistic theories where both signs have solutions and the choice of the sign is a convention. Using the conjugation matrices, the dual representation can be defined as
\begin{equation}
\bar{\psi}_{C}=\psi^T C,\ \ \ \bar{\psi}_{D}=\psi^\dagger D
\end{equation}
where $\psi^\dagger=(\psi^*)^T$ is the complex conjugate of the transposition of $\psi$, and the subscript $C$ and $D$ to denotes charge conjugate and Dirac conjugate.  The Majorana condition is to identify the charge conjugate with the Dirac conjugate which reads,
\begin{equation}\label{Mc12245}
\bar{\psi}_{C}=\bar{\psi}_{D}.
\end{equation}

There are two families of solutions to the defining equation \eqref{gammadef} of the Gamma matrices in two dimensions.
If $\Gamma^1\neq0$, the non-equivalent solutions can be written as
\begin{equation}
 \quad \Gamma^1=\left(\begin{aligned}
0 \ \ & 0\\
2\ \ & 0
\end{aligned}
\right),\ \ \ \ \Gamma^2=\left(\begin{aligned}
1\ & \,\,\,\,\,\,\,0\\
0\ & -1
\end{aligned}
\right),
\label{choiceI}\end{equation}
where we have fixed an overall coefficient in $\Gamma^1$, and an overall sign in $\Gamma^2$.  The later is just a convention, and the former is chosen to make the action of the  boost transformation take the standard form of Jordon cell,
\begin{equation}
M_0\psi=\frac{1}{2}\Gamma^{1}\Gamma^2\psi=\left(
\begin{aligned}
0\ \ & 0\\
1\ \ & 0
\end{aligned}
\right )\left(
\begin{aligned}
& \psi_1\\
& \psi_2
\end{aligned}
\right )\label{M0psi}
\end{equation}
By changing the relative normalization between $\psi_1$ and $\psi_2$, it is always possible to make the above choice. As we will see in next section, the spinor actually form a primary multiplet under the BMS algebra.

As mentioned earlier, a dual spinor can be defined by using either the charge conjugation matrices $C$ or the Dirac conjugation matrix $D$ \eqref{conjugation}.
Due to the fact that the Gamma matrices \eqref{choiceI} are real, the two matrices $C$ and $D$ as defined in \eqref{conjugation} become indistinguishable.
Now the Majorana condition \eqref{Mc12245} becomes simply the reality condition \eq{\psi^*=\psi.\label{realfermion}}
In this paper, we will focus on Majorana fermions, whose dual is given by
\begin{equation}\label{chargeConj}
\bar{\psi}\equiv\bar{\psi}_{C}=\bar{\psi}_{D}=\psi^T C
\end{equation}
To solve $C$, we can substitute the Gamma matrices \eqref{choiceI} into the equation \eqref{conjugation}.
It turns out that only the equation with a minus sign, namely $C\Gamma^a=-(\Gamma^a)^TC$, allows a non-trivial solution. With this choice, the charge conjugation matrix is given by,
\begin{equation}
C=\left(\begin{aligned}
0\ \  & 1\\
-1\ \ & 0
\end{aligned}
\right)
\end{equation}
Thus, we have found the simplest bi-spinor $\bar\psi \psi$ invariant under the action of $M_0$.

To discuss bi-spinors systematically, it is also important to consider how they transform under inverting the ``absolute direction", $x^1\rightarrow -x^1$, which is the  usual parity transformation in Carrollian theory, or time reversal in Galilean theory. We will still refer to this parity transformation for convenience. In terms of components, the bi-spinor $\bar{\psi} \psi  = 2\psi_1\psi_2$ is linear in the absolute direction, and hence is parity odd.
Under the parity transformation, the gamma matrices transform as
\eq{\Gamma^1\rightarrow -\Gamma^1,\ \ \Gamma^2\rightarrow \Gamma^2}
Similar to the $\Gamma^5$ in the usual 4d relativistic Clifford algebra, we can also construct a matrix $\Gamma^*$ which is boost invariant but parity odd,
\eq{
\Gamma^*&={1\over 2}\epsilon_{ab}\Gamma^{a}\Gamma^{b}=\Gamma^1\\
\Gamma^*&\rightarrow -\Gamma^* \quad under \quad x^1\rightarrow -x^1
}
With the help of
$\Gamma^*$, we can construct the following bi-spinors as listed in Table.\ref{bispinors}.
\begin{table}[htp]
\caption{bi-spinors}\label{bispinors}
\begin{center}
\begin{tabular}{|c|c|c|}
\hline
spinors& boost \& parity \\
\hline
$\bar\psi\Gamma^*\psi$&scalar  \\
$\bar\psi\psi$&pseudo-scalar  \\
  $\bar\psi\Gamma^*\Gamma^a\psi$& vector \\
   $\bar\psi\Gamma^a\psi$&pseudo-vector \\
   \hline
\end{tabular}
\end{center}
\label{default}
\end{table}%

Now we have enough ingredients to build the action for fermions.  Let us consider an action in the following form,
\begin{equation}
S=\frac{1}{8\pi g}\int e^a \wedge e^b \epsilon_{ab}\,  \mathcal L
\end{equation}
where $g$ is an overall coefficient to be fixed later and $e^a \wedge e^b \epsilon_{ab}$ is the boost invariant volume form, which has weight $-2$ under the scaling symmetry.
As the action should be invariant under both the boost and the rigid scaling transformation,
the Lagrangian density $\mathcal L$ has to be the boost invariant with scaling weight $2$.
In a fermionic model, the Lagrangian $\mathcal  L$ contains a kinetic term in the form of a  bi-spinor with one derivative.
From the list above, we can build kinetic terms from either the vector or pseudo-vector
\begin{equation}
 \mathcal L_{v}=-i\bar\psi\Gamma^*\Gamma^a e_{a}^{~\mu}D_\mu \psi\ \ \ or\ \  \mathcal L_{pv}=-i\bar \psi\Gamma^a e_{a}^{~\mu}D_\mu\psi
\end{equation}
where $D_\mu$ is the covariant derivative acting on the spinors \eqref{covD}, which becomes $\partial_\mu$ in the flat geometry. As $D_\mu$ has dimension one, the above action will be scaling invariant provided that $\psi$ has scaling weight $1/2$. Note that there is a prefactor $i$ to ensure the action is real.
One may wonder whether it is possible to add a mass term using the scalars or pseudo-scalars. If we wish to keep scaling invariance, mass terms cannot be added because it does not have the desired scaling dimension.
This is different from the WCFT fermion.
Thus, there are two types of free fermions with the Gamma matrices \eqref{choiceI}. In the flat G/C geometry, they can be expressed in terms of the components as
\begin{equation}
S_{v}=-\frac{i}{4\pi g}\int dxdy \,\psi_1\partial_y\psi_1
\end{equation}
and
\begin{equation}\label{bmsfermion}
S_{pv}=-\frac{i}{2\pi g}\int dxdy (\psi_1\partial_x\psi_1-\psi_2\partial_y\psi_1)
\end{equation}
The first one, constructed from the vector in Table.\ref{bispinors}, is a chiral fermion in a CFT$_2$ which has been studied thoroughly in the literature.  In the rest of this paper, we focus on the fermionic model \eqref{bmsfermion} constructed from the pseudo-vector  with Carrollian symmetry, which we refer to as the BMS fermion. The name will become clearer in the next section.
By construction, the action \eqref{bmsfermion} is parity even, and is invariant under  the G/C  boost, rigid scaling, and translational transformations. The action is  real provided that the reality condition \eqref{realfermion} is satisfied, namely $\psi_1$ and $\psi_2$ are both real.\footnote{Our action \eqref{bmsfermion} looks similar with eq. (4.4) of \cite{Bagchi:2017cte} which was obtained from the study of tensionless limit of superstring theory. The action looks the same up to a relative coefficient between the two components of $\psi$.
We note that, however, their conjugate relation eq. (4.8c) indicates that the two components are complex conjugate to each other, whereas we require them to be both real. On the other hand, \cite{Bagchi:2018wsn} worked with the same action with real fermions.}

Finally, let us comment on the other choice of the gamma matrices which satisfy \eqref{gammadef},
\eq{
\quad \Gamma^1= 0,\ \ \ \ (\Gamma^2)^2=I}
A convenient choice is \eq{
\Gamma^{1}=0,\ \ \ \ \Gamma^2=\left(\begin{aligned}
1\ & 0\\
0\ & 1
\end{aligned}
\right),\ \ \ \ C=\left(\begin{aligned}
1\ & 0\\
0\ & 1
\end{aligned}
\right)
\label{chiralgamma}
}
By a similar discussion, we find that in the flat coordinates, free fermions can only be\begin{equation}
S_{chiral}=-{i\over 4\pi g}\int dxdy (\psi_1\partial_y\psi_1+\psi_2\partial_y\psi_2)\label{chiralfermion}
\end{equation}
which is the action of two real chiral fermions.

\subsubsection{Interacting theories }
To construct interacting theories, we can consider two spinors $\psi$ and $\chi$, with the choice of gamma matrices \eqref{choiceI}.
Table \ref{bispinors} allows us to add a four-fermion interaction to the action, so that
\begin{equation}
S=\int e^a \wedge e^b \epsilon_{ab}\Big(-i\bar \psi\Gamma^ae_{a}^{~\mu}\partial_\mu \psi -i\bar \chi \Gamma^ae_{a}^{~\mu}\partial_\mu \chi+\lambda \bar\psi\psi \bar\chi\chi\Big)
\end{equation}
More generally, we can consider $N_1$ BMS fermions  and $N_2$ chiral fermions.  The general  action with four-fermion interactions can be expressed as
\begin{equation}
S\propto\int e^a \wedge e^b \epsilon_{ab}(K-V)
\end{equation}
where $K$ represents the kinematic terms of $N_1$ BMS fermions and $N_2$ chiral fermions,
\begin{equation}
K=-i\sum_{I=1}^{N_1} \bar \psi^I\Gamma^a e_{a}^{~\mu}\partial_\mu\psi^I-i\sum_{I=N_1+1}^{N_1+N_2} \bar \psi^I\Gamma^*\Gamma^a e_{a}^{~\mu}\partial_\mu\psi^I\,,
\end{equation}
and $V$ represents the interaction term,
\begin{equation}
V=\lambda^{IJKL}_1\psi^I_1\psi^J_1\psi^K_1\psi^L_1+\lambda^{IJKL}_2\psi^I_1\psi^J_1\psi^K_1\psi^L_2+\lambda^{IJKL}_3\psi^I_1\psi^J_{2}\psi^K_{1}\psi^L_{2}
\end{equation}
It is interesting to study these interacting theories with G/C and rigid scaling symmetries, which we postpone to future work.

\section{The free BMS fermion model}
We have seen that there are two types of free fermion models with G/C symmetries in the last section, one of which is trivially a holomorphic sector of the usual 2d Majorana fermion, well-studied in the CFT$_2$ literatures while the other is of great interest with novel properties.
In this section, we will focus on the model \eqref{bmsfermion} with Carrollian symmetry, which we refer to as the BMS fermion model hereafter.
We first consider its action and equations of motion on cylinder and plane, and show that the symmetry is enhanced to the infinite dimensional Carollian conformal symmetry, or isomorphically the BMS$_3$ symmetry, in section \ref{BasicPro}.
Then we perform canonical quantization in the highest weight NS and R vacuum in section \ref{quan}. We find all the primary operators and calculate their correlation functions in section \ref{priandcor}.

\subsection{Symmetries}\label{BasicPro}

In the last section, we have found the general free fermion action \eqref{bmsfermion} which is invariant under the G/C symmetry, and rigid scaling symmetry. In this section, we focus on Carrollian theory by specifying $x^1$ as the  spatial direction, and $x^2$ as the temporal direction.
Let us first put the free fermion model on a cylinder parameterized by \((x^1,x^2)=(\sigma,\tau)\) subject to the identification \((\sigma,\tau)\sim(\sigma + 2 \pi,\tau)\).
Then the action \eqref{bmsfermion} becomes,
\begin{align}
S = -\frac{i}{2\pi}\int d\tau d\sigma (\psi_1\partial_{\sigma}\psi_1-\psi_2\partial_{\tau}\psi_1)\label{BMSf}
\end{align}
where \(\psi_{1},\ \psi_{2}\) are Grassmann variables. In this model, the conjugate momenta with respect to \(\psi_{1}\) and \(\psi_{2}\) can be calculated by the graded Leibnitz rule
\begin{align}
\Pi_{1} &= \frac{\delta L}{\delta (\partial_{\tau}\psi_{1})}=\frac{i}{2\pi} \psi_{2},\\
\Pi_{2} &=\frac{\delta L}{\delta (\partial_{\tau}\psi_{2})}= \frac{i}{2\pi} \psi_{1}.
\end{align}
The equal time (i.e. equal \(\tau\)) anti-commutator between these fermions is then
\begin{align}\label{anti}
\{\psi_{1}(\tau,\sigma),\psi_{2}(\tau,\sigma')\} &=2\pi  \delta(\sigma'-\sigma),\\
 \{\psi_{1}(\tau,\sigma),\psi_{1}(\tau,\sigma')\}&=\{\psi_{2}(\tau,\sigma),\psi_{2}(\tau,\sigma')\}=0.
\end{align}
By construction, the action \eqref{BMSf} is invariant under translations, rigid boost, and rigid scaling. Now we show that the symmetry is actually much larger, and derive it by symmetry enhancements \cite{Chen:2019hbj} together with the minimal input given by the covariant arguments in the last section.
Consider the following infinitesimal translation
\begin{align}
\sigma\rightarrow \sigma + \varepsilon,\quad \tau\rightarrow\tau+\tilde{\varepsilon}
\end{align}
where \(\varepsilon\) and \(\tilde{\varepsilon}\) are constants. We can apply the Noether theorem to obtain two conserved currents,
\begin{align}\label{cur}
2\pi \pmb{j}_{\varepsilon} &= - \varepsilon T   d\sigma  + \varepsilon M d\tau,\\
2\pi \pmb{j}_{\tilde{\varepsilon}} &= -\tilde{\varepsilon}M  d\sigma,
\end{align}
where \(T\) and \(M\) play the role of the stress tensors, and are given by
\begin{align}\label{TM}
T &= -\frac{i}{2}\psi_{1}\p_{\sigma}\psi_{2}-\frac{i}{2}\psi_{2}\p_{\sigma}\psi_{1},\\
M &=-\frac{i}{2}\, \psi_{1}\p_{\tau}\psi_{2}=-{i\over 2} \p_\tau P,\quad P\equiv \psi_1 \psi_2.
\end{align}
Using the equations of motion, it is not difficult to verify that  the current 1-forms \eqref{cur} are closed on-shell, or equivalently the stress tensor satisfies the conservation law,
\begin{align}\label{TMc}
\p_{\tau}(\varepsilon T ) &= \p_{\sigma}(\varepsilon M),\\
\p_{\tau}(\tilde{\varepsilon}M)&=0.\nonumber
\end{align}
The conservation law is readily generalized to more general transformations by the replacement \(\varepsilon\rightarrow\varepsilon(\sigma)\) and \(\tilde{\varepsilon}\rightarrow\tilde{\varepsilon}(\sigma)\).
First we note that the second equation in \eqref{TMc} still holds under the replacement,
while the first equation in \eqref{TMc} will no longer hold, due to the appearance of an extra term $\varepsilon'(\sigma)M$ on the right hand side. The aforementioned extra term, however,  can be compensated if we add a term to the left hand side, so that we get the following conservation relations for two arbitrary functions $\varepsilon(\sigma)$ and $\tilde{\varepsilon}(\sigma)$,
\begin{align}\label{TMBMSc}
\p_{\tau}\Big(\varepsilon(\sigma) T + \varepsilon'(\sigma) \tau M\Big) &= \p_{\sigma}\Big(\varepsilon(\sigma) M\Big),\\
\p_{\tau}\Big(\tilde{\varepsilon}(\sigma)M\Big)&=0.
\end{align}
The above relations suggest new conservation laws. To understand the underlying symmetry, we need to construct the conserved currents and conserved charges, and derive the transformation rules by acting the charges on the fields.
The conservation laws \eqref{TMBMSc} imply that the following currents are on-shell closed \footnote{
In general, the derivation of the Noether currents is ambiguous, and in this case one can still add an exact form to \eqref{curbms}. For instance, we can define currents
\eq{\nonumber
2\pi \pmb{j}_{\varepsilon(\sigma)} &= -\Big( \varepsilon(\sigma) (T+\alpha \p_\sigma P) + (1-2\alpha)\varepsilon'(\sigma) \tau M \Big) d\sigma  + (1-2\alpha)\varepsilon(\sigma) M d\tau,
}
The above modification of the currents  correspond to a modification of $T$, together with a rescaling of $M$. As we will show later,  our choices of $T$ and $M$ are compatible with BMS symmetry.
 },
\begin{align}\label{curbms}
2\pi \pmb{j}_{\varepsilon(\sigma)} &= -\big( \varepsilon(\sigma) T + \varepsilon'(\sigma) \tau M \big) d\sigma  + \varepsilon(\sigma) M d\tau,\\
2\pi \pmb{j}_{\tilde{\varepsilon}(\sigma)} &= -\big( \tilde{\varepsilon}(\sigma) M \big) d\sigma,
\end{align}
conserved charges on the spatial circle with constant $\tau$ are given by
\begin{align}\label{classicalcharges}
Q_{\varepsilon(\sigma)} &=  \int_{\sigma\text{-cycle}} \pmb{j}_{\varepsilon(\sigma)} 
= -\frac{1}{2\pi}\int_0^{2\pi} d\sigma ~\varepsilon(\sigma) \big( T - \tau \p_\sigma M \big),\nonumber\\
Q_{\tilde{\varepsilon}(\sigma)} &= \int_{\sigma\text{-cycle}} \pmb{j}_{\tilde{\varepsilon}(\sigma)} = -\frac{1}{2\pi} \int_0^{2\pi} d\sigma  ~\tilde{\varepsilon}(\sigma) M.
\end{align}
where we have used integration by parts in the first line.
The charges are independent of the time $\tau$, as a consequence of the conservation laws \eqref{TMBMSc} and periodic boundary conditions.
These charges generate the following infinitesimal transformations,
\begin{align}
\delta_{\varepsilon}\psi_{1}(\sigma,\tau) &=\{Q_{\varepsilon},\psi_{1}(\sigma,\tau)\}_{PB}= -\varepsilon \p_{\sigma}\psi_{1} - \frac{1}{2}\varepsilon'\psi_{1},\nonumber\\
\delta_{\varepsilon}\psi_{2}(\sigma,\tau)&=\{Q_{\varepsilon},\psi_{2}(\sigma,\tau)\}_{PB} = -\varepsilon \p_{\sigma}\psi_{2}-2\, \varepsilon'\tau\p_{\sigma}\psi_{1} - \frac{1}{2}\varepsilon'\psi_{2}-\tau\varepsilon''\psi_{1},\\
\delta_{\tilde{\varepsilon}}\psi_{1}(\sigma,\tau)&=\{Q_{\tilde{\varepsilon}},\psi_{1}(\sigma,\tau)\}_{PB}= 0,\nonumber\\
\delta_{\tilde{\varepsilon}}\psi_{2}(\sigma,\tau)&=\{Q_{\tilde{\varepsilon}},\psi_{2}(\sigma,\tau)\}_{PB} = -2\tilde{\varepsilon}\p_{\sigma}\psi_{1} - \tilde{\varepsilon}'\psi_{1}.\nonumber
\end{align}
where we have dropped the dependence on $\sigma$ in $\varepsilon(\sigma)$ and $\tilde{\varepsilon}(\sigma)$ for brevity. They are just the infinitesimal version of the transformation law \cite{Hao:2021urq, Chen:2020vvn} of the rank-2 primary multiplet with weight \(\Delta=\frac{1}{2}\) and boost charge ${\boldsymbol{ \xi}} = \begin{pmatrix}\label{BstChar}
 0&0  \\
 1&0
\end{pmatrix},$
\begin{align}\label{TranRule}
\tilde{\psi}_{1}(\tilde{\tau},\tilde{\sigma}) &= \left|f'\right|^{-\frac{1}{2}}\psi_{1}(\tau,\sigma)\\
\tilde{\psi}_{2}(\tilde{\tau},\tilde{\sigma}) &= \left|f'\right|^{-\frac{1}{2}}\psi_{2}(\tau,\sigma) - \left|f'\right|^{-\frac{3}{2}}\left(\tau f''+g'\right)\psi_{1}(\tau,\sigma).\nonumber
\end{align}
under the BMS transformation
\begin{equation}\label{BMStr}
\sigma\rightarrow f(\sigma),\ \ \tau\rightarrow f'(\sigma)\tau+g(\sigma)
\end{equation}
As a consistency check, one can directly verify that the action \eqref{BMSf} is indeed invariant under the BMS transformation \eqref{BMStr}  with the transformation rule \eqref{TranRule} .

By expanding the parameters $\varepsilon(\sigma), \tilde{\varepsilon}(\sigma)$ in terms of the Fourier modes,
\eq{\label{jchar}
\varepsilon_n = \tilde{\varepsilon}_n =e^{in\sigma}.
}
we obtain infinitely many symmetry generators, which are mode expansions of the conserved charge operators \eq{\label{cylindergen}
L_{n} &\coloneqq  Q_{\varepsilon_n}  \quad M_{n} \coloneqq Q_{\tilde{\varepsilon}_n},}
It is not difficult to verify that they form the BMS algebra under the anti-commutation relation \eqref{anti}.
We will show in section \ref{quan} that the BMS algebra has a central extension after choosing a vacuum and considering normal ordering explicitly.

\subsection{Canonical quantization}\label{quan}
In this subsection we discuss the canonical quantization by choosing  the highest weight vacuum.
Another choice of the vacuum, the induced vacuum, will be discussed in section 6.

The action \eqref{BMSf} on the cylinder should be periodic in $\sigma$,  requiring the boundary conditions on the fundamental fields $\psi_1,\psi_2$ to be either periodic or anti-periodic, referred to as the R and NS sector respectively.
The map from cylinder to plane
\begin{align}\label{cylinder2plane}
x=e^{i\sigma},\quad y=ie^{i\sigma}\tau,
\end{align} allows us to consider the theory on the plane, with the action given by,
\begin{align}\label{bmsplane}
S=\frac{1}{2\pi} \int dxdy (\psi_1\partial_x\psi_1-\psi_2\partial_y\psi_1).
\end{align}
For convenience, from now on we will mainly carry out the calculations on the plane, unless  stated otherwise.
Due to the fact that the fermions $\psi_1,\psi_2$ have conformal weight \(\frac{1}{2}\),  the periodic/anti-periodic boundary conditions on the cylinder become anti-periodic/periodic on the plane under the cylinder-to-plane map, so that on the plane we have
\begin{align}
\psi_{i}(e^{2\pi i} x) &= +\psi_{i}(x),\quad  \text{Neveu–Schwarz sector (NS)},\quad i=1,2\\
\psi_{i}(e^{2\pi i} x) &= -\psi_{i}(x),\quad \text{Ramond sector (R)},
\end{align}
The equations of motion are
\begin{align}\label{EoM}
\partial_{y}\psi_{1} &= 0,\\
\partial_{y}\psi_{2} &= 2\,\partial_{x}\psi_{1},
\end{align}
with the following solution in terms of the Laurent expansion,
\begin{align}\label{ABmode}
\psi_1=\sum\limits_{n}{\defa}_{n}x^{-n-\frac{1}{2}} ,\quad\psi_2=\sum\limits_{n}{\defb}_{n}x^{-n-\frac{1}{2}}+2y\partial_x \psi_1,
\end{align}
where \(n \in \Z\) for the R sector, \(n\in \Z +\frac{1}{2}\) for the NS sector. The reality condition \eqref{realfermion} then implies, \begin{equation}
{\defa}_n^\dagger={\defa}_{-n},\ {\defb}_n^\dagger={\defb}_{-n}\label{reality}
\end{equation}
Now canonical quantization can be carried out on the plane with the following anti-commutation relation
\begin{align}\label{CommRela}
\{\psi_{1}(x,y),\psi_{2}(x',y)\} =2\pi \delta(x'-x).
\end{align}
which can be equivalently written in terms of the mode operators
\begin{equation}\label{c347r}
\{{\defa}_n,{\defb}_m\}=\delta_{n+m,0},\quad
\{{\defa}_n,\defa_m\}=\{{\defb}_n,{\defb}_m\}=0.
\end{equation}
The anti-commutation relations \eqref{c347r} are valid on both the cylinder and the plane.

Taking the cylinder to plane map and the analytical continuation into account, the quantum version of the classical Noether currents \eqref{TM} whose corresponding charges generate translations along $x$ and $y$ now become operators,
\begin{align}
T=-\frac{1}{2}:\psi_2\partial_x\psi_1:-\frac{1}{2}:\psi_1\partial_x\psi_2:,\ \ M=-:\psi_1\partial_x\psi_1:,
\end{align}
where the definition of the normal ordering $:\cdots:$ depends on the choice of the vacuum, to be specified momentarily. Here we would like to keep the normal ordering implicit. The currents can be expanded in Laurent series as
\eq{
T&=\sum_{n} L_n x^{-n-2}-\sum_n (n+1)yM_{n-1}x^{-n-2},  \label{T_modes}\\
M&=\sum_n M_n x^{-n-2}, \label{M_modes}
}
which can be inverted to define infinitely many charges $L_n$ and $M_n$. Using the reality condition \eqref{realfermion}, the Hermitian conjugates are given by
\eq{L_n^\dagger=L_{-n},\quad M_n^\dagger=M_{-n}.} The charges are the quantum version of the classical charges \eqref{cylindergen} on the plane.
Using the transformation law \eqref{TranRule} under the plane-to-cylinder map \eqref{cylinder2plane}, the zero-mode generator of the Virasoro algebra on the cylinder has a shift,
$L_0^{cyl}=L_0^{pl} -\frac{1}{12}.
$ We focus on the plane in this paper.

\subsubsection*{The highest weight NS vacuum}

In the NS sector, modes are labeled by half integers, without any zero modes.  This leads to a natural choice for the vacuum,
\begin{align}\label{NSVac}
{\defa}_{n}|0\rangle &= 0, \quad n\geq \frac{1}{2}\nonumber\\
{\defb}_{n}|0\rangle &= 0, \quad n\geq \frac{1}{2}
\end{align}
and a prescription for the normal ordering
\begin{align}
:{\defa}_{n}{\defb}_{m}: = \left\{\begin{array}{lll}
\quad {\defa}_{n}{\defb}_{m} \qquad n\leq -\frac{1}{2}\\
\\
-{\defb}_{m}{\defa}_{n} \qquad n \geq  \frac{1}{2}\\
\end{array}\right.\label{normalordering}
\end{align}
Inverting the relations \eqref{T_modes} and \eqref{M_modes}, and plugging in mode expansion, we find  the following expression for the symmetry generators on the plane
\begin{align}\label{LMs}
L_{n} &= \sum\limits_{k} \left(k-\frac{n}{2}\right):{\defa}_{n-k}{\defb}_{k}:\, \\
M_{n} &= \sum\limits_{k} \left(k+\frac{1}{2}\right):{\defa}_{n-k}{\defa}_{k}: \nonumber \end{align}
where \eq{k\in \mathbb Z+{1\over2} } as we are in the NS sector.
Further using the commutation relations \eqref{c347r}, we find that the charges
\eqref{LMs} form a centrally extended BMS algebra,
\begin{align}\label{bmsalg}
[L_{n},L_{m}] &= (n-m)L_{m+n} + \frac{1}{12}(n^3-n)\delta_{n+m,0},\nonumber\\
[L_{n},M_{m}] &= (n-m)M_{m+n},\\
[M_{n},M_{m}] &= 0\nonumber
\end{align}
where the central charges are given by  \(c_{L} = 1\) and \(c_{M} = 0\).
The BMS algebra contains a sub-algebra $\mathfrak{iso}(2,1)$ with generators $\{L_{0,\pm1},M_{0,\pm1}\}$, which is often referred to as the global BMS algebra. As we will see later, the NS vacuum is invariant under the global BMS algebra, whereas the R sector vacuum is not.

Now it is straightforward to check that the vacuum \eqref{NSVac} is a highest weight vacuum of the BMS algebra, namely
\begin{align}
L_{n}|0\rangle &= 0,\qquad n\geq -1\\
M_{n}|0\rangle &= 0,\qquad n\geq -1
\end{align}\leavevmode
In particular, the generators of the global sub-algebra of the BMS$_3$ algebra all annihilate the vacuum.

Let $\vec{i}\equiv (i_1,\,i_2\,\cdots),\, \vec{j}\equiv (j_1,\,j_2\,\cdots)$, then the state space in the NS sector  is spanned by
\begin{equation}\label{NSbasis}
|\vec{i},\vec{j}\rangle :={\defa}_{-1/2}^{i_1}{\defa}_{-3/2}^{i_2}\cdots {\defb}_{-1/2}^{j_1}{\defb}_{-3/2}^{j_2}\cdots|0\rangle \quad i_{n},j_{m}=0,1.
\end{equation}

\subsubsection*{The highest weight R vacuum}
In the R sector, modes are labeled by integers which include zero.
Similar to CFTs, the zero modes should be considered separately. The algebra of the zero modes is given by
\begin{equation}\label{algzero}
\{\beta_{0},\beta_{0}\}=\{\gamma_{0},\gamma_{0}\}=0,\ \ \ \{\beta_{0},\gamma_{0}\}=1
\end{equation}
This suggests that the vacuum in the R sector is degenerate.
To further characterize the vacuum, we can define a target-space spin operator,
\eq{
S\equiv \gamma_0\beta_0-{1\over2}
}
whose eigenstate can be raised and lowered by $\beta_0$ and $\gamma_0$, namely \eq{  [S,\beta_0]=-\beta_0, \quad  [S,\gamma_0]=\gamma_0}
This suggests that the vacuum in the R sector is a doublet of the target-space spin operator, labeled by the eigenvalue $|s\rangle_R$, with
\eq{
S|s\rangle_R=s|s\rangle_R,\quad s=\pm {1\over2}
}
Therefore, the vacua in the R sector $|s\rangle_R$ are related to each other by the action of $\beta_0$ or $\gamma_0$, and annihilated by positive-integer modes, namely
\begin{align}\label{Rvac}
\beta_{0}\left|\frac{1}{2}\right\rangle& = \left|-\frac{1}{2}\right\rangle ,\ \ \ \beta_{0}\left|-\frac{1}{2}\right\rangle =0\\
\gamma_{0}\left|-\frac{1}{2}\right\rangle& = \left|\frac{1}{2}\right\rangle ,\ \ \ \gamma_{0}\left|\frac{1}{2}\right\rangle =0\nonumber\\
{\defa}_{n}\left|s\right\rangle_R& =  0, \quad n> 0,\quad n\in \mathbb{Z}\nonumber\\
{\defb}_{n}\left|s\right\rangle_R &= 0, \nonumber
\end{align}
Due to the degeneracy of the vacua in the R sector, zero modes should be treated separately.
The normal ordered product for other modes reads
\begin{align}
:{\defa}_{n}{\defb}_{m}: = \left\{\begin{array}{lll}
\quad {\defa}_{n}{\defb}_{m} \qquad n\leq -1\\
\\
-{\defb}_{m}{\defa}_{n} \qquad n \geq  1\\
\end{array}\right.
\end{align}
The symmetry generators, with the exception of $L_0$, can all be written as \eqref{LMs}, but now with
\eq{k\in \mathbb Z.}
The generator $L_0$ contains zero modes, whose ordering can be traced back to the original definition from \eqref{T_modes},
\begin{equation}
\begin{split}
L_{0}=&\sum_{k\in\mathbb{Z},k\neq 0}k:\beta_{k}\gamma_{-k}:+\frac{1}{8}(\beta_{0}\gamma_{0}+\gamma_{0}\beta_{0})\\
=&\sum_{k\in\mathbb{Z},k\neq 0}k:\beta_{k}\gamma_{-k}:+\frac{1}{8}\label{RL0}
\end{split}
\end{equation}
Acting on the ground state, we thus have
\begin{equation}
\begin{split}
L_{0}\left|s\right\rangle_{R}=\frac{1}{8}\left|s\right\rangle_{R},\ \ \ \ M_{0}\left|s\right\rangle_{R}=0\label{L0charge}
\end{split}
\end{equation}
where $|s\rangle_{R}$ denotes the vacuum doublet, $|s\rangle_{R}=(\left|\frac{1}{2}\right\rangle, \,\left|-\frac{1}{2}\right\rangle)^T$. Although the vacua in the R sector is degenerate, each of the vacuum states is a singlet of the BMS algebra with weight $\frac{1}{8}$, and boost charge $\xi=0$. In this case, the definition indeed satisfies the highest weight condition,
\begin{align}
L_{n}|s\rangle_R = 0,\qquad M_{n}|s\rangle_R = 0,\qquad n\geq 1\label{RvacLMn}
\end{align}\leavevmode
In contrast to the case in the NS sector, we note that the vacuum in the R sector is not invariant under the global part of the BMS algebra. We have seen that they carry non-vanishing $L_0$ charge \eqref{L0charge}. In addition, the vacuum is not translational invariant either, as
\begin{equation}\label{Rtrans}
L_{-1}|s\rangle_R=\frac{1}{2}(\beta_{-1}\gamma_0-\beta_0\gamma_{-1})|s\rangle_R,\ \ M_{-1}|s\rangle_R=\beta_{-1}\beta_0|s\rangle_R
\end{equation}
We can also check that the generators $L_n$ and $M_n$ form the same BMS algebra with central charge \(c_{L} = 1\) and \(c_{M} = 0\).

 Let $\vec{i}\equiv (i_1,\,i_2\,\cdots),\, \vec{j}\equiv (j_1,\,j_2\,\cdots)$, then the state space is spanned by
\begin{equation}\label{basis}
|\vec{i},\vec{j},s\rangle :={\defa}_{-1}^{i_1}{\defa}_{-2}^{i_2}\cdots {\defb}_{-1}^{j_1}{\defb}_{-2}^{j_2}\cdots|s\rangle_{R}\quad i_{n},j_{m}=0,1,\ \ s=\pm {1\over2}.
\end{equation}

\subsection{Primary operators and correlation functions}\label{priandcor}
In this subsection we calculate the propogators of the fundamental fermion field, find all the primary operators, and compute their correlation functions.

\subsubsection*{General results for BMSFT}
As discussed in \cite{Hao:2021urq,Chen:2020vvn,Chen:2022jhx},
BMS field theories feature multiplets, on which the action of $L_0$ is diagonal while the action of $M_0$ is block diagonal, consisting of Jordan blocks.
For a multiplet $\bO\equiv (O_1,\cdots O_r)^T$ with rank $r$,
let  $\Delta$ denotes the conformal weight which is the eigenvalue of $L_0$,
  ${{\bxi}}$ denotes the boost charge matrix which is
   a $r\times r$ Jordan cell with diagonal element $\xi$.
Then the defining property for a highest weight multiplet is  that the OPEs with the stress tensors $T$ and $M$ take the following form,
\begin{align}\label{PriMul}
T(\tilde{x},\tilde{y}) \bO(x,y) & \sim \frac{\Delta \bO}{(\tilde{x}-x)^2}- \frac{2(\tilde{y}-y) {\bxi} \bO}{(\tilde{x}-x)^3}+\frac{\partial_x \bO}{\tilde{x}-x} -\frac{(\tilde{y}-y)\partial_y \bO}{(\tilde{x}-x)^2} ,\nonumber \\
M(\tilde{x},\tilde{y}) \bO(x,y) & \sim  \frac{{\bxi} \bO}{(\tilde{x}-x)^2}+\frac{\partial_y \bO}{\tilde{x}-x} .
\end{align}
This can be used to find all the primary operators in the theory.
The general form of two and three point functions for the BMS highest weight multiplets with rank \(r\) read
\eq{
&\langle O_{ia}(x_1,y_1)O_{jb}(x_2,y_2)\rangle =\left\{\begin{array}{ll}
0& \mbox{for $q<0$}\\
\, \delta_{ij}\,  d_r\,  |x_{12}|^{-2\Delta_i} e^{-2\xi_i\frac{y_{12}}{x_{12}}}\frac{1}{q!}\left(-\frac{2y_{12}}{x_{12}}\right)^q,& \mbox{otherwise},\end{array} \right.\label{BMS2pf}\\
&\langle O_{ia}(x_1,y_1)O_{jb}(x_2,y_2)O_{kc}(x_3,y_3)\rangle =A \,B\, C_{ijk}\label{3pf}
}
where \(q=a+b+1-r\); $i,\,j$ and $k$ label the multiplets, while $a,\,b$ and $c$ label the components within a multiplet, and
\begin{eqnarray}
A&=&\exp(-\xi_{123}\frac{y_{12}}{x_{12}}-\xi_{312}\frac{y_{31}}{x_{31}}-\xi_{231}\frac{y_{23}}{x_{23}}),\\
B&=&|x_{12}|^{-\Delta_{123}}|x_{23}|^{-\Delta_{231}}|x_{31}|^{-\Delta_{312}},\\
C_{ijk;abc}&=&\sum_{n_1=0}^{a-1}\sum_{n_2=0}^{b-1}\sum_{n_3=0}^{c-1}c_{ijk}^{(n_1 n_2n_3)}\frac{(p_i)^{a-1-n_1}(p_j)^{b-1-n_2}(p_k)^{c-1-n_3}}{(a-1-n_1)!(b-1-n_2)!(c-1-n_3)!},
\end{eqnarray}
with
\begin{equation}
p_i=\partial_{\xi_i}\ln A.
\end{equation}

\subsubsection*{Propogators and OPEs}
Now let us turn to our BMS fermion model. For both choices of the vacuum \eqref{NSVac} and \eqref{Rvac},  the propogator can be defined as,
\eq{\label{2ptdef}
&\langle O_1(x_1,y_1)O_2(x_2,y_2)\rangle\\
=&\langle 0|\mathcal{X}(O_1(x_1,y_1)O_2(x_2,y_2))|0\rangle-\langle 0|:O_1 (x_1,y_1)O_2(x_2,y_2):|0\rangle
\nonumber}
where $:\cdots:$ denotes the normal ordering compatible with the specific choice of the vacuum,
 and $\mathcal{X}(\cdots)$ denotes the radial ordering on the complexified $x$-plane, the latter of which is further related to the time ordering on the Lorentz cylinder, as explained in \cite{Hao:2021urq}.
 Taking into account the fermionic nature of the fields,  the radial ordering can be defined as,
\begin{equation}
\mathcal{X}\left(\psi_{\alpha}(x_{1})\psi_{\beta}(x_{2})\right) = \left\{\begin{array}{lll}
+\psi_{\alpha}(x_{1}) \psi_{\beta}(x_{2}) & \text { for } & |x_{1}|>|x_{2}| \\
-\psi_{\beta}(x_{2}) \psi_{\alpha}(x_{1}) & \text { for } & |x_{1}|<|x_{2}| \\
\end{array}\right.
\end{equation}
where $a,b=1,2$.
Further using the mode expansion \eqref{ABmode}, we can calculate the propogators in the highest NS vacuum,
\begin{align}\label{362NSp}
\langle\psi_{1}(x_{1})\psi_{1}(x_{2})\rangle_{NS} &= 0\nonumber\\
\langle\psi_{1}(x_{1})\psi_{2}(x_{2},y_{2}) \rangle_{NS} &= \frac{1}{x_{1}-x_{2}}\\
\langle\psi_{2}(x_{1},y_{1})\psi_{2}(x_{2},y_{2}) \rangle_{NS} &= -\frac{2(y_{1}-y_{2})}{(x_{1}-x_{2})^{2}},\nonumber
\end{align}
which take the general form of \eqref{BMS2pf} for a rank-2 multiplet with $\Delta={1\over2}$ and $\xi=0$. In the highest weight R vacuum,  the propogators are given by
\begin{align}
\langle\psi_{1}(x_{1})\psi_{1}(x_{2})\rangle_R &= 0\nonumber
\\
\langle\psi_{1}(x_{1})\psi_{2}(x_{2},y_{2})\rangle_{R}&=\frac{x_1+x_2}{2\sqrt{x_1x_2}(x_1-x_2)}
\\
\langle\psi_{2}(x_{1},y_{1})\psi_{2}(x_{2},y_{2}) \rangle_{R}&=\frac{x_{2}y_{1}\left(-x_1^2-4x_{1}x_{2}+x_{2}^2\right) +x_{1}y_{2} \left(-x_1^2+4x_{1}x_{2}+x_{2}^2\right)}{2\left(x_1-x_2\right)^2\left(x_1x_2\right)^{3/2}}\nonumber
\end{align}
which are not in the general form of \eqref{BMS2pf}. In particular, the propogators in the R vacuum are not invariant under translations. This is due to the fact that the R vacua is not invariant under the global part of the BMS algebra \eqref{Rtrans}.  As we will show later, the R sector vaccua can be understood as inserting a twist operator at the origin of the NS vacuum, so that all correlators in the R sector can be described in terms of correlators in the NS sector. Therefore, it is enough to discuss correlators in the NS sector.

When one operator approaches the other, the correlation functions in the R sector agree with the ones in the NS sector and therefore they have the same OPEs in both case, whose leading terms reads,
\begin{align}\label{BasOPE}
\psi_{1}(x_{1})\psi_{1}(x_{2})&\sim 0+\cdots,\nonumber\\
\psi_{1}(x_{1})\psi_{2}(x_{2},y_{2}) &\sim \frac{1}{x_{1}-x_{2}}+\cdots,\nonumber\\
\psi_{2}(x_{1},y_{1})\psi_{2}(x_{2},y_{2}) &\sim -\frac{2(y_{1}-y_{2})}{(x_{1}-x_{2})^{2}}+\cdots.
\end{align}
The OPEs of other operators can then be obtained from \eqref{BasOPE} via Wick contractions. In particular, we note that the OPEs among the stress tensors read,
\eq{\label{TMope}
T(x',y') T(x,y) & \sim \frac{1}{2(x'-x)^4}+ \frac{2T(x,y)}{(x'-x)^2}- \frac{4(y'-y) M(x,y)}{(x'-x)^3} \\
& +\frac{\partial_x T(x,y)}{x'-x}-\frac{(y'-y)\partial_y T(x,y)}{(x'-x)^2} ,\nonumber\\
T(x',y') M(x,y) & \sim \frac{2M(x,y)}{(x'-x)^2}+\frac{\partial_x M(x,y)}{x'-x},\nonumber\\
M(x',y') M(x,y) & \sim 0.\nonumber
}
which takes the standard form of OPEs among stress tensors and is consistent with the BMS algebra \eqref{bmsalg}.
From the most singular terms, we again read the central charges $c_L=1$ and $c_M=0$.

\subsubsection{The BMS data in the NS sector}\label{NSdata}
Let us now determine the BMS data of this free fermion model by finding the primary operators and calculating the three-point coefficients. By calculating the OPEs with the stress tensors and comparing with the standard form \eqref{PriMul}, we can find the following primary fields,  \begin{itemize}
\item The identity operator $I$ with $\Delta=\xi=0$.
\item The fundamental fermions $\bpsi=(\psi_1,\psi_2)^T$  form a highest weight multiplet with rank $r=2$,
with conformal weight $1/2$ and boost charge $0$, or written in terms of matrix
\begin{equation}\label{weightpsi}\Delta=\left(\begin{matrix}
   \frac{1}{2}&0  \\
   0&\frac{1}{2}\end{matrix}\right),\ \
\bxi=\left(\begin{matrix}
   0&0  \\
   1&0\end{matrix}\right).
\end{equation}
The propogators \eqref{362NSp} are consistent with this.
\item The composite operator $P\equiv:\psi_1\psi_2:$ is a primary operator of rank \(1\), with $\Delta=1,\,\xi=0$. Due to the fermionic nature of the fields in this model,  $P$ is the only composite operator without derivatives. We will see that there are no more primary operators in the next subsection.
\end{itemize}
To describe this model in terms of BMS data, we still need the three-point coefficients among $\psi_1,\,\psi_2,\,P$. We then find
that the three point functions indeed take the general form \eqref{3pf}, with
 non-vanishing coefficients
\begin{equation}\label{pope}
c_{\bpsi\bpsi P}^{(010)}=1,\ \ c_{\bpsi\bpsi P}^{(100)}=-1
\end{equation}

\subsection{The operator basis and state operator correspondence}
Now let us find a basis of local operators in the model,
which can be built from the fundamental fields $\psi_1,\psi_2$, and their derivatives with respect to the coordinates $x$ and $y$. From the equations of motion \eqref{EoM}, we learn that  $\p_y\psi_1$ does not appear in a local operator. Moreover, the action of $\partial_y$ on $\psi_2$ can always be replaced by acting on $\partial_x$ on $\psi_1$. This means that the operator basis can be chosen as
\begin{equation}
\{\prod_{k,m=0}^{\infty}(\partial_x^k\psi_1)^{i_n}(\partial_x^m\psi_2)^{j_m}\},\ \ i_n,j_m=0,1.
\end{equation}
where each order of derivative on $\psi_\alpha$ can appear at most once due to its fermionic nature.
It is straightforward to see that the only primary operators are $I,\psi_1,\psi_2,P=:\psi_1\psi_2:$.

In the NS sector, state-operator correspondence can be built up by inserting the local operators at the origin.
Using the mode expansion \eqref{ABmode}, we have
\begin{equation}
\begin{split}
\lim_{x\rightarrow 0,y\rightarrow 0}\partial_{x}^{k}\psi_{1}\left|0\right\rangle=&k!{\defa}_{-k-\frac{1}{2}}\left|0\right\rangle\\
\lim_{x\rightarrow 0,y\rightarrow 0}\partial_{x}^{k}\psi_{2}\left|0\right\rangle=&k!{\defb}_{-k-\frac{1}{2}}\left|0\right\rangle.
\end{split}
\end{equation}
so that we have state-operator correspondence
 \eq{\label{stateoperator}
:\partial_x^{k}\psi_1\,  \partial_x^{m}\psi_2\,\,\cdots:\sim\,\,  :{\defa}_{-k-\frac{1}{2}}\,{\defb}_{-m-\frac{1}{2}}\,\cdots:|0\rangle}
The right hand side is precisely the Fock states in the NS sector \eqref{NSbasis}.

\subsubsection*{Twist operators for the R vacua}

Now we consider the R sector, which can be equivalently described by interpreting the R sector vacuum as inserting a primary operator at the origin of the NS sector vacuum,
\begin{equation}
\left|\frac{1}{2}\right\rangle\equiv\sigma|0\rangle,\ \ \ \ \left|-\frac{1}{2}\right\rangle\equiv\mu|0\rangle
\end{equation}
where we have omitted the subscripts $R$ and $NS$ for simplicity.
From the properties of the vacuum states \eqref{L0charge} and \eqref{RvacLMn}, we learn that the twist operators are both primary operators with conformal weight
$\Delta=\frac{1}{8}$ and $\xi=0$. Using the general result of two point functions \eqref{BMS2pf},  we can thus get the following correlators
\begin{equation}
\langle \sigma\sigma\rangle=\langle \mu\mu\rangle=\frac{1}{x_{12}^{\frac{1}{4}}},\ \ \langle \sigma\mu\rangle=0
\end{equation}
Note that  the twist operators cannot be built from the fundamental field $\psi_{1},\psi_{2}$.
The relation between the two vacuum states \eqref{Rvac} corresponds to the following  OPEs
\begin{equation}\label{twsope}
\begin{split}
\psi_{1}(x',y')\sigma(x,y)\sim& \frac{\mu(x,y)}{(x'-x)^{\frac{1}{2}}},\\
\psi_{2}(x',y')\mu(x,y)\sim& \frac{\sigma(x,y)}{(x'-x)^{\frac{1}{2}}}
\end{split}
\end{equation}
Any state $|O\rangle_R$ in the R sector can be obtained by inserting the composite operator $:O\sigma:$ or  $:O\mu:$ to the origin of the NS vacuum.

Together with the BMS data in the NS sector \eqref{NSdata},  we have the following fusion algebra,
\begin{equation}\label{fusalg}
\begin{split}
[\bpsi]\times [\sigma]\rightarrow [\mu],\ \ \ & [\bpsi]\times[\mu]\rightarrow[\sigma],\ \ \ [\sigma]\times[\mu]\rightarrow[\bpsi]\\
[P]\times[P]\rightarrow[I],\ \ \ & [\sigma]\times [\sigma]\rightarrow [I],\ \ \ [\mu]\times [\mu]\rightarrow [I]\\
 [\bpsi]\times [P]\rightarrow [\bpsi], \ \ \   & [\bpsi]\times [\bpsi]\rightarrow [I]+ [P]
\end{split}
\end{equation}
where $[O]$ refers to the BMS highest weight module represented by the primary operator $O$, which can be either singlet or multiplet. The first line of \eqref{fusalg} is from the defining property of the R-vacua \eqref{twsope}, the second line is from the vanishing three-point coefficients containing the operators $P,\sigma,\mu$, and the third line is from the non-vanishing three-point coefficients \eqref{pope}.

\section{The staggered module}
In this section we will begin with a general definition of the so-called staggered module, and then show that states in the BMS fermion module can be organized into BMS staggered modules.
We have organized the state space in terms of the modes ${\defa}_n,{\defb}_n$ in the last section. Now we want to find how it carries the representation of the BMS algebra. To do so, we consider the action of $L_n,M_m$ on the states in section \ref{section4.1}.

CFT$_2$s can be organized into the  highest weight representation of Virasora algebra, which means that the state space can be decomposed into different highest weight modules represented by the primary states.
One may expect that a BMS invariant theory can be similarly organized into BMS modules.
However, in the BMS scalar model \cite{Hao:2021urq}, there exists at least one state which is neither a primary state itself nor a descendant state of the BMS algebra. To accommodate this type of operator, the ordinary highest weight module should be enlarged to the so-called staggered module, which is a reducible but indecomposable representation of the BMS algebra, defined as the semi-direct sum of ordinary highest weight representations.

In \cite{Rohsiepe:1996qj,Kytola:2009ax}, the staggered module $\mathcal{S}$ for Virasoro algebra can be defined via the following short exact sequence,
\begin{equation}\label{short}
0\longrightarrow \mathcal{H}^L\stackrel{\iota}{\longrightarrow}\mathcal{S}\stackrel{\pi}{\longrightarrow}\mathcal{H}^R\longrightarrow0
\end{equation}
where $\mathcal{H}^L$ and $\mathcal{H}^R$ are irreducible highest weight modules, also named as typical modules \cite{Kytola:2009ax}, $\iota$ and $\pi$ are module homomorphisms. There is also a central element $Q$ acting non-diagonalizably on the highest weight vectors, possessing Jordan cells of rank-2. In other words, the staggered module $\mathcal{S}$ has a submodule isomorphic to a typical highest weight module $\mathcal{H}^L$, and the quotient $\mathcal{S}/\mathcal{H}^L$ is ismorphic to another typical highest weight module $\mathcal{H}^R$. As a consistence requirement, the central charges of $\mathcal{H}^L$ and $\mathcal{H}^R$ must coincide. In logarithmic CFTs \cite{Kytola:2009ax}, the central element $Q$ is $L_0$.
The above construction can be generalized by replacing the building block in \eqref{short}, namely the typical modules, by any given representation of any symmetry group.
For the BMS scalar, the symmetry algebra is BMS instead of Virasoro, and we can choose the central element as $Q=M_0$. Starting from a highest weight representation $\mathcal{S}^{(1)}$,
we can build a staggered model $\mathcal S^{(2)}$ by replacing the typical modules $\mathcal H^L$ and $\mathcal H^R$ in the short sequence \eqref{short}
by $\mathcal S^{(1)}$ and $\mathcal S^{(2)}/\mathcal S^{(1)}$, respectively. Then we can further use $\mathcal S^{(2)}$ as building blocks to construct another short sequence, so that we can obtain another staggered module $\mathcal S^{(3)}$. Similarly, we can construct $\mathcal S^{(n+1)}$ from an extension of $\mathcal S^{(n)}$,
\begin{equation}
0\longrightarrow \mathcal{S}^{(n)}\stackrel{\iota}{\longrightarrow}\mathcal{S}^{( n+1)}\stackrel{\pi}{\longrightarrow}\mathcal{S}^{(n+1)}/\mathcal{S}^{(n)}\longrightarrow0
\end{equation}
 This procedure can be carried out  successively and we can even have a staggered module with $n\to \infty$,
with a Hilbert space
\begin{equation}
{\mathcal H}=\mathcal{S}^{(1)}\oplus_S\mathcal{S}^{(2)}/\mathcal{S}^{(1)}\oplus_S\cdots
\end{equation}
The BMS scalar model \cite{Hao:2021urq} can be organized into such a BMS staggered module $\mathcal S^{(\infty)}$ with  $\mathcal S^{(1)}$ chosen as  irreducible highest weight modules, while as  we will see later that the BMS fermion has the same structure  $\mathcal S^{(\infty)}$ but $\mathcal S^{(1)}$ can also be chosen as reducible highest weight modules.

\subsection{Enlarged staggered module in the NS sector}\label{section4.1}
Now we discuss how to organize states in the BMS fermion model \eqref{bmsplane}.
For computational convenience, it is useful to write down the commutation relations between $L_n,M_n$ and ${\defa}_n,{\defb}_n$, which read,
\begin{align}\label{basic}
[L_n,\defa_m]&=-\left(\frac{n}{2}+m\right){\defa}_{n+m},\quad [L_n,{\defb}_m]=-\left(\frac{n}{2}+m\right){\defb}_{n+m}.\\
[M_n,\defa_m]&=0,\ \ [M_n,{\defb}_m]=-(n+2m){\defa}_{n+m}.\nonumber
\end{align}
From the above expression, we note that
\begin{equation}
[L_0,{\defa}_{n_1}\cdots {\defb}_{m_1}\cdots]=-(n_1+\cdots+m_1+\cdots){\defa}_{n_1}\cdots {\defb}_{m_1}\cdots
\end{equation}
which is valid for both the NS sector and the R sector. As discussed in the last section, states in the R sector can be understood as states in the NS sector dressed by the twist operator. In the following discussion, we focus on the NS sector.
\begin{itemize}
\item At weight $0$, there is the vacuum state$|0\rangle$.
\item At weight $\frac{1}{2}$, there are two states,
\begin{equation}
\psi_1\sim {\defa}_{-\frac{1}{2}}|0\rangle,\ \ \ \psi_2\sim {\defb}_{-\frac{1}{2}}|0\rangle,
\end{equation}
which are primary states with $\Delta={1\over2}$ and $\xi=0$, consistent with the OPE result \eqref{weightpsi}.
\item
At weight $1$, there is only one state created by
\begin{equation}
P\sim {\defa}_{-\frac{1}{2}}{\defb}_{-\frac{1}{2}}|0\rangle
\end{equation}
which is a primary singlet.
\item At weight $\frac{3}{2}$, there are two states created by
\begin{equation}
\partial_x\psi_1\sim{\defa}_{-\frac{3}{2}}|0\rangle=L_{-1}{\defa}_{-\frac{1}{2}}|0\rangle ,\ \ \partial_x\psi_2\sim{\defb}_{-\frac{3}{2}}|0\rangle=L_{-1}{\defb}_{-\frac{1}{2}}|0\rangle
\end{equation}
which are descendant states.
\item
At weight $2$,  there are four states
\end{itemize}
\eq{\label{MTDK}
M&\sim-{\defa}_{-\frac{1}{2}}{\defa}_{-\frac{3}{2}}|0\rangle,\nonumber \\
 T&\sim -\frac{1}{2}(  {\defa}_{-\frac{1}{2}}{\defb}_{-\frac{3}{2}} -{\defa}_{-\frac{3}{2}}{\defb}_{-\frac{1}{2}}  )|0\rangle,\quad
 \p_x P\sim\Big( {\defa}_{-\frac{1}{2}}{\defb}_{-\frac{3}{2}}+ {\defa}_{-\frac{3}{2}}{\defb}_{-\frac{1}{2}}\Big)|0\rangle \\
K&\sim  -{1\over4}  {\defb}_{-\frac{1}{2}}{\defb}_{-\frac{3}{2}}|0\rangle\nonumber
}
where a new operator  $K$ shows up.
The action of \(M_0\) on the basis of
\begin{align}
\Big(|2M\rangle,\,|T\rangle,\,|K\rangle,\,|\p_{x}P\rangle=L_{-1}|P\rangle \Big)
\end{align}
is given
\begin{equation}
\left(\begin{matrix}
   0&0&0&0  \\
   1&0&0&0\\
   0&1&0&\frac{1}{4}\\
   -1&0&0&0\\
   \end{matrix}\right).
\end{equation}
which is not in the form of standard Jordan blocks, due to a mixing between the vacuum BMS module and the $P=:\psi_1\psi_2:$ module.
To better understand these states, we check the action of BMS annihilation generators $L_n, \,M_n,\,n>0$. The non-vanishing terms are
\eq{\label{LnMnp}
L_2|T\rangle&=L_2L_{-2}|0\rangle=\frac{1}{2}|0\rangle\nonumber\\
L_1|\p_x P\rangle&=L_1(L_{-1}|P\rangle)=2|P\rangle\\
M_1|K\rangle&={1\over 2}|P\rangle, \quad M_2|K\rangle=\frac{1}{4}|0\rangle,\ \ \ \nonumber
}
First $|\p_xP\rangle=L_{-1}|P\rangle$ is a descendant state of the primary state $|P\rangle$.
The states $(|2M\rangle, \,|T\rangle)$ are invariant under the action of $L_1$ and $M_1$, the annihilation operators of global BMS generators, and thus form a quasi-primary multiplet with $\Delta=2, \,\xi=0$.
The new operator $K$ is neither a BMS primary state  nor a descendant, and thus requires an extension of the highest weight representation. $K$ is mapped to the primary state $|P\rangle$ by the action of $M_1$, and thus is not a quasi-primary either.
This is in contrast to the case for the BMS scalar, where $(2M, \,T, \,K)$ form a quasi-primary multiplet with rank $3$.

As a side remark, we note that the  action  of \(M_0\) can be brought to the standard Jordan block by a generalized diagonalization procedure,  \begin{equation}
\left(\begin{matrix}
   0&0&0&0\\
   1&0&0&0\\
   0&1&0&0\\
   0&0&0&0\\
   \end{matrix}\right)
\end{equation}
with a new basis corresponding to the operators
\eq{\left(|2M\rangle, \frac{4}{3}|T\rangle+\frac{1}{3}|\p_{x}P\rangle,\frac{4}{3}|K\rangle,2|T\rangle+2|\p_{x}P\rangle\right),}
Despite this simple structure of $M_0$, it turns out this basis is not convenient for organizing the states. Because of the mixing of $|T\rangle$ with the descendant state $|\p_x P\rangle$, the first two states does not form a quasi-primary multiplet.
We will focus on the basis \eqref{MTDK} in subsequent discussions.

Finally, let us calculate the OPEs between the four operators at level two.
From the previous discussion, the two components $(2M,T)$ in the basis form a rank-2 multiplet and the OPEs are already derived in \eqref{TMope}. The OPEs of the remaining basis operators $K,D$ can be calculated by Wick theorem
\begin{equation}\label{TKOPE}
\begin{split}
T(x,y)K(0,0)&\sim-\frac{y}{x^{5}}-\frac{3yP}{2x^{4}}-\frac{2yT}{x^{3}}-\frac{y\p_xP}{2x^{3}}+\frac{2K}{x^{2}}+\frac{1}{x}\partial_{x}K-\frac{y}{x^{2}}\partial_{y}K\\
M(x,y)K(0,0)&\sim\frac{1}{4x^{4}}+\frac{P}{2x^{3}}+\frac{T}{x^{2}}+\frac{\p_xP}{4x^{2}}+\frac{1}{x}\partial_{y}K\\
T(x,y)\p_xP(0,0)&\sim\frac{2P}{x^{3}}+\frac{4yM}{x^{3}}+\frac{2\p_xP}{x^{2}}-\frac{y}{x^{2}}\partial_{y}\p_xP+\frac{1}{x}\partial_{x}\p_xP\\
M(x,y)\p_xP(0,0)&\sim\frac{-2M}{x^{2}}+\frac{1}{x}\partial_{y}\p_xP
\end{split}
\end{equation}
One can check that the above OPEs can be alternatively obtained from the operator state correspondence.

\subsubsection*{The combined module of $I$ and $P$}
Let us first consider the vacuum module.
Due to the mixing between the stress tensor and $\p_xP$, we have to consider the identity module and the $P$ module together.
Using the relations \eqref{LnMnp}, we can draw the following diagram for the states with integer weights up to level $\Delta=2$,
\begin{figure}[H]\centering
\begin{tikzcd}[scale cd = 1.2, column sep=small]
\Delta=0&{\color{black}\times}&&{\color{black}\bullet}\arrow[lldd, blue]\arrow[dd, rightharpoonup, bend left=3, black]&&&&{\color{black}\times}\\
\Delta=1&&&&&{\color{black}\bullet}\arrow[lllld,blue]\arrow[d, rightharpoonup, bend left=3, black]&&&\\
\Delta=2&{\color{blue}\bullet}&&{\color{black}\bullet}\arrow[ll,blue] \arrow[uu, rightharpoonup, bend left=3, black]\arrow[uu, rightharpoonup, bend right=3, black]\arrow[lluu, dashed,blue]&\ &{\color{black}\bullet}\arrow[llll,bend left=20,blue]\arrow[u, rightharpoonup, bend left=3, black]&&{\color{red}\bullet}\arrow[lll, bend right=30,blue]\arrow[lllluu,bend right=20, blue]\arrow[uu, dashed]\arrow[llu, blue]
\end{tikzcd}
\end{figure}
\noindent where we use $\times $ for null states, the black dot with $\Delta=0$ for the vacuum,  the black dot with $\Delta=1$ for $|P\rangle$, the blue dot for $|M\rangle$,  the two black dots at $\Delta=2$ for $|T\rangle $ and $|\p_x P\rangle$ respectively, and the red dot  for  $|K\rangle$. The blue/black arrows  represent  the action of $M_n$/$L_n$, with upwards/downwards arrow for positive/negative $n$. The action of $M_0$ runs horizontally, and is always toward the left.
The arrow ending in the middle of two dots represents their linear combination.

Now let us organize these in the language of BMS staggered model we introduced earlier in this section.
The key step is to correctly identify the building block $\mathcal S^{(1)}$. For the BMS scalar \cite{Hao:2021urq}, $\mathcal S^{(1)}$ was chosen to be the irreducible highest weight modules. For the BMS fermion, however, this is not possible for the vacuum module, due to the mixing between descendants of the identity $I$ and another primary $P$.
Therefore, the minimal build block can be chosen to be the reducible highest weight module $\mathcal{H}_{I,P}$, which consists of
 two BMS primary states, the vacuum and $|P\rangle$, as well as their BMS descendant states.
The module $\mathcal{H}_{I,P}$ contains two submodule  $\mathcal{H}_I$ and $\mathcal{H}_P$, each of which is an irreducible highest weight module. It is not a direct sum, due to the relation $M=-{1\over2}\p_y P$, or equivalently due to the existence of the null state
\begin{equation}
\mathcal{N}\equiv  2M_{-2}|0\rangle+M_{-1}|P\rangle\sim 0
\end{equation}
Modding out the null state mixes states in the two submodules $\mathcal{H}_I$ and $\mathcal{H}_P$.
As a result, the building block for the staggered module will be
\begin{equation}
\mathcal{S}^{(1)}\equiv \mathcal{H}_{I,P}=\frac{\mathcal{H}_I\oplus\mathcal{H}_P}{\mathcal H_{\mathcal{N}}}
\end{equation}
where  $\mathcal H_{\mathcal{N}}$ is spanned by the null state $\mathcal{N}$.
The module $\mathcal{H}_{I,P}$ is represented by the following diagram,
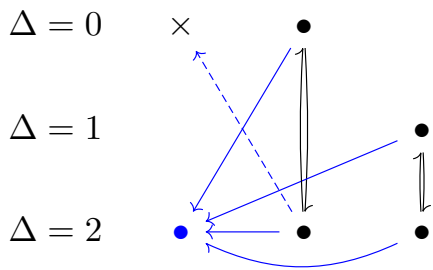
\begin{figure}[H]\centering
\begin{tikzcd}[scale cd = 1.2, column sep=small]
\Delta=0&{\color{black}\times}&&{\color{black}\bullet}\arrow[lldd, blue]\arrow[dd, rightharpoonup, bend left=3, black]&&&\\
\Delta=1&&&&&{\color{black}\bullet}\arrow[lllld, blue]\arrow[d, rightharpoonup, bend left=3, black]\\
\Delta=2&{\color{blue}\bullet}&&{\color{black}\bullet}\arrow[uu, rightharpoonup, bend left=3, black]\arrow[ll, blue]\arrow[lluu, dashed,blue]&&{\color{black}\bullet}\arrow[u, rightharpoonup, bend left=3, black]\arrow[llll, rightharpoonup, bend left=25, blue]\\
\end{tikzcd} \label{IP}
\end{figure}
Now let us use the additional operator $K$ to seed more states by considering the composite operators using $K$s.
As a first step, we can construct the composite operators with the primaries, which in the case of $\mathcal{H}_{I,P}$ is $K$ and $:KP:$.
By acting on $I$, $P$, $K$ and $:KP:$ with the BMS generators, we can build an enlarged highest weight module $\mathcal{S}^{(2)}$. After moding out $\mathcal{H}_{I,P}$, the new states $K$ and $:KP:$ become the BMS primary states, so that the quotient $\mathcal{H}_{KI,KP}\equiv \mathcal{S}^{(2)}/\mathcal{S}^{(1)}$ is isomorphic to $\mathcal{H}_{I,P}$ itself.
This means that we indeed have the short exact sequence \begin{equation}
0\longrightarrow \mathcal{H}_{I,P}\stackrel{\iota}{\longrightarrow}\mathcal{S}^{(2)}\stackrel{\pi}{\longrightarrow}\mathcal{H}_{KI,KP}\longrightarrow0
\end{equation}
Equivalently, we can write $\mathcal{S}^{(2)}$ as a semi-direct sum
\begin{equation}
\mathcal{S}^{(2)}=\mathcal{H}_{I,P}\oplus_S\mathcal{H}_{KI,KP}
\end{equation}
Similarly, we can build $\mathcal{S}^{(3)}$ by adding new seeds $:KKI:$ and $:KKP:$ to $\mathcal{S}^{(2)}$, and moreover build $\mathcal{S}^{(n)}$ by adding  new seeds $:K^{n-1}I:$ and $:K^{n-1}P:$ to $\mathcal{S}^{(n-1)}$, namely
\begin{equation}
\mathcal{S}^{(3)}=\mathcal{S}^{(2)}\oplus_S\mathcal{H}_{KKI,KKP}, \ \ \ \mathcal{S}^{(n)}=\mathcal{S}^{(n-1)}\oplus_S\mathcal{H}_{K^{n-1}I,K^{n-1}P}
\end{equation}
Therefore the full staggered module that includes the vacuum and the primary $P$ is a semi-direct sum of infinite many highest weight modules, each of which is isomorphic to the starting building block $\mathcal{H}_{I,P}$, and the seed operator for the sum is the operator $K$.
\begin{equation}
\mathcal{S}_{I,P}=\mathcal{H}_{I,P}\oplus_S\mathcal{H}_{KI,KP}\oplus_S\cdots\oplus_S\mathcal{H}_{K^{n}I,K^{n}P}\oplus_S\cdots
\end{equation}

\subsubsection*{The $\boldsymbol{\psi}$ module}

Now let us consider the $\boldsymbol{\psi}=(\psi_{1},\psi_{2})$ module. We list the first a few states below according to their conformal weights,
\begin{itemize}
\item $\Delta={1\over2}:\ |\psi_{1}\rangle,\,|\psi_{2}\rangle$
\item $\Delta={3\over2}: \ L_{-1}|\psi_{1}\rangle,\ L_{-1}|\psi_{2}\rangle$
\item $\Delta=\frac{5}{2}: \ L^{2}_{-1}|\psi_{1}\rangle,\ L^{2}_{-1}|\psi_{2}\rangle$;
$M_{-2}|\psi_{2}\rangle\ L_{-2}|\psi_{1}\rangle,\ L_{-2}|\psi_{2}\rangle,\ |K\psi_{1}\rangle.$
\end{itemize}
The two states at $\Delta={1\over2}$ form a primary multiplet with rank $2$, acting $L_{-n},\,M_{-n},n>0$ on which generates a BMS highest weight module $H_{\bpsi}$.
The state $|K\psi_1\rangle$ with $\Delta={5\over2}$ is neither a primary nor descendant, and hence will seed a staggered module.
Analogous to the previous discussion, we choose $\mathcal{S}^{(1)}=H_{\bpsi}$, and  build $\mathcal{S}^{(2)}$ by considering the composite operator $:K\bpsi:$, the later of which contains a null state $:K\psi_2:$. Thus only $:K\psi_1:$ will seed the enlarged module $\mathcal{S}^{(2)}$.
After modding out $H_{\bpsi}$, we have $\mathcal{S}^{(2)}/\mathcal{S}^{(1)}=\mathcal{H}_{K\psi_1}$, where $\mathcal{H}_{K\psi_1}$ is an irreducible BMS highest module generated by $K\psi_1$.
 Then we have the following short exact sequence,
\begin{equation}
0\longrightarrow \mathcal{H}_{\bpsi}\stackrel{\iota}{\longrightarrow}\mathcal{S}^{(2)}\stackrel{\pi}{\longrightarrow}\mathcal{H}_{K\psi_1}\longrightarrow0
\end{equation}
Due to the fact that $:K\psi_2:$ is null, the quotient $\mathcal{S}^{(2)}/\mathcal{S}^{(1)}=\mathcal{H}_{K\psi_1}$ is not isomorphic to $\mathcal{H}_{\bpsi}$. This is different from the vacuum module, where the quotient is also isomorphic to the initial building block $\mathcal{S}^{(1)}$.
Carrying out the above procedure successively,
we get \eq{\mathcal{S}^{(n)}=\mathcal{S}^{(n-1)}\oplus_S\mathcal{H}_{K^{n-1}\psi_1},}
so that the entire $\bpsi$ module can be written as a semi-direct sum of different BMS highest weight modules represented by $\bpsi,\,:K\psi_1:, \,:K^2\psi_1:,\cdots $, namely
\begin{equation}
\mathcal{S}_{\bpsi}=\mathcal{H}_{\bpsi}\oplus_S\mathcal{H}_{K\psi_1}\oplus_S\cdots\oplus_S\mathcal{H}_{K^{n}\psi_1}\oplus_S\cdots
\end{equation}

\section{Torus partition function}
In this section, we consider the torus partition function of the BMS fermion \eqref{BMSf}. We first review the modular invariance of the BMSFT following \cite{Jiang:2017ecm} and then calculate the torus partition function of the BMS fermion in the highest weight vacuum explicitly.
Torus partition in the induced vacuum is carried out in section 6.
\subsection{Modular Property of the BMSFTs}
We first give a quick review on the torus partition function in the BMSFTs here. We consider a torus which is determined by two identifications on a two dimensional plane,
\eq{
(canonical) \,spatial\, circle:\quad &(\tau,\sigma)\sim(\tau,\sigma+2\pi)\label{canonical}\\
thermal\, circle:\quad &(\tau,\sigma)\sim(\tau-2\pi i b,\sigma-2\pi i a)
}
It is useful to embed $\mathbb R^2$ into $\mathbb C^2$ in the subsequent discussions. The partition function on the above torus is formally a path integral over all fields satisfying boundary conditions specified by the two identifications.
Alternatively, the torus partition function can be written as a trace over the state space which is determined by the {\it spatial circle}, weighted by the evolution along the {\it thermal circle},
\begin{equation}\label{zbth}
Z(a, b)=Tr\,e^{-2\pi a(L_0-\frac{c_L}{24})-2\pi b (M_0-\frac{c_M}{24})}
\end{equation}
where the translational generators are defined on the cylinder with the spatial circle \eqref{canonical}, which we refer to as the canonical circle.
More generally, a torus can be described by the fundamental region on the plane
\begin{equation}
(\tau,\,\sigma)\sim(\tau,\,\sigma)+m\,\vec{\beta}_S +n\,\vec{\beta}_T
\end{equation}
where $m$ and $n$ are integers, so that the torus is completely determined by a pair of vectors $\vec{\beta}_S,\, \vec{\beta}_T$ on the plane.
For instance, the torus \eqref{canonical} has a canonical spatial circle $\vec{\beta}_S=(0,\,2\pi)$, and a thermal circle $\vec{\beta}_T
=(-2\pi i b,\, -2\pi i a)$.
The transformations acting on the plane that leave the torus invariant form the modular group, $SL(2,\,\mathbb{Z})/\mathbb{Z}_2$.  The action of $SL(2,\,\mathbb{Z})$ is given by
\begin{equation}
\left(
\begin{aligned}
a\ \ &b\\
c\ \ &d
\end{aligned}
\right)\left(
\begin{aligned}
\vec{\beta}_T\\
\vec{\beta}_S
\end{aligned}
\right)=
\left(
\begin{aligned}
\vec{\beta}_T'\\
\vec{\beta}_S'\end{aligned}
\right)
\end{equation}
with
\begin{equation}
ad-bc=1, \ \ \ \ a,b,c,d\in \mathbb{Z}.
\end{equation} The reason to mod $\mathbb{Z}_2$ is because the simultaneous inversion of all the matrix elements does not change the torus.
The modular group is generated by the $T$ and $S$ transformations, with \eq{
T=\left( \ba{cc}
1&1\\
0&1
\ea
\right),\quad
S=\left( \ba{cc}
0&-1\\
1&0
\ea
\right).}
The S-transformation corresponds to swapping of spatial circle and the thermal circle, followed by a scaling which brings the new spatial circle to have period $2\pi$. As a result, the partition function has to satisfy
\begin{equation}
(\tau,\sigma)\sim\left(\tau+\frac{2\pi ib}{a^2},\sigma-\frac{2\pi i}{a}\right)\sim(\tau,\sigma-2\pi)
\end{equation}
The S-invariance of the partition function is then
\eq{
Z(a,b)=Z\left(\frac{1}{a},-\frac{b}{a^2}\right).\label{modularinvS}
}
The T-transformation adds the spatial circle to the thermal circle, giving a new identification,
\begin{equation}
(\tau,\sigma)\sim(\tau,\sigma+2\pi)\sim(\tau-2\pi i b,\sigma-2\pi i a+2\pi)
\end{equation}
If the torus partition function has modular T-invariance,   it has to satisfy
\eq{
Z(a,b)=Z(a+i,b).\label{modularinvT}
}
All the other transformations in the $SL(2, \mathbb{Z})$ can be obtained by group multiplication of the $T-$ and $S-$ transformations.
The $U\equiv TST$ transformation is of particular interest, the action of which on the torus is provided here for completeness,
\begin{align}
(\tau,\sigma)\sim (\tau,\sigma+2\pi)\sim \left(\tau-\frac{2\pi ib}{(1-ia)^2},\sigma-\frac{2\pi ia}{1-ia}\right),
\end{align}
Then the U invariance of the partition function is to require
\begin{align}
Z(a,b) = Z\left(\frac{a}{1-ia},-\frac{b}{(1-ia)^2}\right).
\end{align}

\subsection{Partition function for BMS free fermion}
In this subsection we calculate the partition function for the BMS fermion model on a torus with a spatial circle and a thermal circle as in \eqref{canonical}. Boundary conditions along the spatial circle can be conveniently parameterized by a parameter $\mu$, with $\mu=0$ for the R sector, and $\mu={1\over2}$ for the NS sector. Similarly, the thermal circle also admits either periodic (R) or anti-periodic (NS) boundary conditions, labeled by $\nu=0$ or $\nu={1\over2}$ respectively. Due to the fermionic nature, imposing the R boundary conditions can be realized by inserting $(-1)^F$ into the torus partition function, where \(F\) is the fermion number operator.
Altogether there are four combinations in the choices of  boundary conditions,  labeled by the pair $(\mu,\nu)$,
\eq{\label{ptf}
Z_{\mu,\nu}(a,b)=Tr_{(\mu)}(-1)^{(1-2\nu)F}e^{-2\pi a(L_{0}-\frac{1}{24})-2\pi bM_0}
}
where the trace depends on the boundary conditions along the spatial circle, which we now specify. The state space in the NS vacuum is spanned by \eqref{NSbasis}
\begin{equation}
|\vec{i},\vec{j}\rangle={\defa}_{-1/2}^{i_1}{\defa}_{-3/2}^{i_2}\cdots {\defb}_{-1/2}^{j_1}{\defb}_{-3/2}^{j_2}\cdots|0\rangle.
\end{equation}
Using the conjugation relation \eqref{reality}, we can construct a dual basis
\begin{equation}\label{dualbasis}
\langle\vec{i},\vec{j}|=\langle 0|{\defb}_{1/2}^{j_1}{\defb}_{3/2}^{j_2}\cdots {\defa}_{1/2}^{i_1}{\defa}_{3/2}^{i_2}\cdots .
\end{equation}
The basis is not orthonormal as can be seen from the following inner products,
\begin{equation}\label{inner}
\langle \vec{i'},\vec{j'}|\vec{i},\vec{j}\rangle= N_{NS;\vec{i}\vec{j},\vec{i'}\vec{j'}}\,,\quad
N_{NS;\vec{i}\vec{j},\vec{i'}\vec{j'}}=\delta_{\vec{i'},\vec{j}}\delta_{\vec{j'},\vec{i}}
\end{equation}
Similar to the scalar model, it is convenient to introduce the dual orthonormal basis,
\eq{
&
^{\ \ \ \vee}\langle \vec{i},\vec{j}| \equiv    \sum_{\{\vec{i'},\vec{j'}\}} (N_{NS}^{-1})_{\vec{i}\vec{j},\vec{i'}\vec{j'}} \langle \vec{i'},\vec{j'}|,\ \
^{\ \ \ \vee}\langle \vec{i},\vec{j}| \vec{i'}\vec{j'}\rangle=\delta_{\vec{i},\vec{j}; \vec{i'},\vec{j'}}.
}
where $N_{NS}^{-1} $ denotes the matrix inverse of  $N_{NS}$ whose matrix elements are defined in \eqref{inner}. Now the trace over the states in the NS sector reads,
\begin{equation}
Tr_{NS}=\sum_{\vec{i},\vec{j}}\,|\vec{i},\vec{j}\rangle\ ^{ \ \vee}\langle \vec{i},\vec{j}|.
\end{equation}
Similar discussions can be carried out in  the R sector with the highest weight vacuum, with the states \eqref{basis},
\begin{equation}
|\vec{i},\vec{j},s\rangle :={\defa}_{-1}^{i_1}{\defa}_{-2}^{i_2}\cdots {\defb}_{-1}^{j_1}{\defb}_{-2}^{j_2}\cdots|s\rangle\quad i_{n},j_{m}=0,1.
\end{equation}
and the dual orthonormal basis can be defined accordingly. The trace over the R sector then reads,
\begin{equation}
Tr_{R}=\sum_{\vec{i},\vec{j},s}\,|\vec{i},\vec{j},s\rangle\ \ ^{\ \vee}\langle \vec{i},\vec{j},s|.
\end{equation}
Note that
\begin{equation}\label{m0action}
^\vee\langle\vec{i},\vec{j}|M_0|\vec{i},\vec{j}\rangle=0,\quad ^\vee\langle\vec{i},\vec{j},s|M_0|\vec{i},\vec{j},s\rangle=0
\end{equation}
namely, the expectation value of $M_0$ vanishes for both the NS and R sector and therefore does not play any role in the calculation of the torus partition function.
As a result, the calculation of the torus partition function \eqref{ptf} amounts to counting the  spectrum of $L_0$, and we get
\begin{equation}
Z_{\mu,\nu}(a,b)=\frac{\theta_{1/2-\mu,1/2-\nu}(ia)}{\eta(ia)},\ \ \theta_{\mu,\nu}(ia)=\sum_{n\in Z}q^{\frac{1}{2}(n+\mu)^2e^{2\pi i(n+\mu)\nu}},
\end{equation}
Under the T and S transformation, the modular properties for the Jacobi theta function are
\eq{\theta_{\mu,\nu}(t+1)&=e^{-i\pi \mu(\nu-1)}\theta_{\mu,\mu+\nu-1/2}(t)
\\
\theta_{\mu,\nu}(-\frac{1}{t})&=\sqrt{-i t}e^{2\pi i \mu \nu}\theta_{\nu,-\mu}(t)
}
namely the $T$ and $S$ transformations exchange $Z_{\mu,\nu}$ among themselves.
It is then straight forward to check that the total partition function is modular $S$ invariant,
\eq{\label{bmspartition}Z(a,b)\equiv \sum_{\mu,\nu=0,{1\over2}} Z_{\mu,\nu} (a,b)=Z\left({1\over a},-{b\over a^2}\right) }  which indeed satisfy \eqref{modularinvS}.
For the $T$ transformation, however, additional phase factors show up and the total partition function is no longer $T$ invariant. It is still possible to make the torus partition function modular invariant under the full $SL(2,\mathbb {Z})$ including both the $T$ and the $S$ transformations by taking 24 copies of the original theory.

To end this section, we compare the partition function of the BMS fermion model  \eqref{bmspartition} with that of two chiral fermions \eqref{chiralfermion}. The agreement is due to the fact that the torus partition function of the BMS fermions is determined only by the $L_0$ spectrum of the theory, which is the same as that of the chiral fermion theory. However, this does not indicate that they are the same theory. We have seen explicitly that the BMS fermion has a non-trivial $y$ dependence and the staggered module, while the chiral fermion has no $y$ dependence and the module is the Virasoro highest weight module.

\section{The induced representation}
In this section, we will consider the induced vacuum of the BMS fermion \eqref{bmsfermion}. We solve the model in the induced vacuum and compute the correlation functions in section 6.1 and then consider the state space and torus partition function in section 6.2 and 6.3.

\subsection{The induced representation revisited}
In this subsection we provide a general discussion on the induced representation for the BMS group. We introduce the notions of primary, descendants, and multiplets in the induced presentation, which are analogous to the discussions in the highest weight representation.

As a BMSFT can be obtained from the UR limit of a CFT$_2$, it is natural to take  the UR limit of  the highest weight representation in CFT$_2$. As was discussed in \cite{Bagchi:2017cpu}, this procedure leads to an indecomposible representation of the BMS group induced by a representation of its Abelian ideal generated by $M_n$s. Such a representation is called an induced representation \cite{Barnich:2014kra,Barnich:2015uva}.

Let us first consider a primary operator in CFT$_2$, satisfying
\eq{\label{cftpri}
[L_0^+,\, O]=h O,  \quad [L_0^-,\, O]=\bar h O, \\
[L_n^+,\, O]=[L_n^-,\, O]=0,\quad \forall n>0\nonumber
}
where $L_n^+,L_n^-$ denotes the Virasoro generators in CFT$_2$. Under the UR limit, the two copies of Virasoro algebras become the BMS algebra via the Wigner-In\"{o}n\"{u} contraction \cite{contraction},
\eq{\ \ L_n=L^+_{n}-L^-_{-n},\ \ M_n=\epsilon(L^+_{n}+L^-_{-n}),
}
so that the primary field \eqref{cftpri} satisfies the induced condition,
\eq{
[L_0,O]&=\Delta O,\qquad [M_0,O]=\xi O\nonumber, \\
 [M_n,O]&=0,\,\ \ \quad\quad\, \forall n\neq0, \,n\in \mathbb{Z}.\label{induced}
}
In particular, the induced vacuum denoted by $|0_I\rangle$ should satisfy
\begin{equation}\label{i72v}
L_0|0_I\rangle=M_n|0_I\rangle=0,\ \ \quad\quad \forall n\in \mathbb{Z}.
\end{equation}

Note that the condition \eqref{induced} is not for all states in the induced representation.
By construction, the operator $O$ originated from a primary in the CFT$_2$ before the UR limit, and therefore still plays the role of a primary field in the BMSFT. From now on, we will refer to an operator satisfying conditions \eqref{induced} as a primary operator, or a primary singlet to be distinguished from the primary multiplet to be introduced momentarily.
We can construct descendants by acting $L_n,\,\forall n\in \mathbb Z$ on the primary $O$.
The simplest descendant is \eq{[L_n,O]\equiv {O}_{(n)}}
Using Jacobi identity, one can easily verify that
\eq{[L_0,\,{O}_{(n)}]&=\Delta_{(n)} \,{O}_{(n)} , \quad \Delta_{(n)}=\Delta-n ,\\
[M_0,\,{O}_{(n)}]&=\xi \,{O}_{(n)},\\
 [M_m,\,{O}_{(n)}]&=2\,m\,\xi\, O \,\delta_{m+n}}
which means that $\,{O}_{(n)}$ is  i)  an eigenstate of $L_0$ with eigenvalue $\Delta-n$,
ii) an eigenstate of $M_0$ with eigenvalue $\xi$,
and iii)  no longer annihilated by all the $M_m$s with $m\neq0$.
More general descendant states can be obtained by acting on $O$ with multiple $L_n$s, and one can verify that the aforementioned three properties are still satisfied by the descendants constructed this way.
One exceptional case in the above discussion is when $\xi=0$, where the action of $L_n$ will bring a primary singlet to another primary singlet with $\xi=0$. They form a representation of one copy of Virasoro algebra only, with the eigenvalue of $L_0$ unbounded.

Similar to the highest weight representation, the primary field in the induced representation may also be a multiplet,  satisfying
\eq{
&[L_0,{\bf O}]= \Delta {\bf O}\nonumber\\
&[M_0,{\bf O}]={ \bxi} \bf O,\label{inducedm} \\
&[M_n,{\bf O}]=0,\,\ \ \quad\quad\, \forall n\neq0, \,n\in \mathbb{Z}.\nonumber
}
where now ${\bxi}$ is a Jordan cell with diagonal element $\xi$.
Similar to the previous discussion on the singlet, descendants can also be generated by acting on the primary multiplet $\bf O$ by $L_n$s with $n\neq0, \,n\in\mathbb Z$.

To summarize, in the induced representation, a BMS module contains a primary operator satisfying \eqref{induced} or a primary multiplet satisfying \eqref{inducedm}, and descendants obtained from the primary singlet or multiplet by acting with $L_n$s. The descendants do not satisfy the conditions in the last line of \eqref{induced} and \eqref{inducedm}.
We will see in section 6.2 that this is the case in the free BMS fermion model.

\subsection{Correlators}
In the NS sector, the induced vacuum can be chosen as the state invariant under the global BMS group, similarly to the highest weight vacuum discussed in section 3.
In terms of the modes, the induced vacuum condition in the NS sector reads
\begin{align}
{\defa}_n|0_I\rangle_{NS}=0,\ \  \forall n\in \mathbb{Z}+\frac{1}{2}
\end{align}
which means that all the $\beta_n$ modes are annihilation operators.
One can directly verify that
\begin{equation}\label{trivial}
L_n|0_I\rangle_{NS}=M_n|0_I\rangle_{NS}=0,\ \ \forall n\in \mathbb{Z}.
\end{equation}
Namely, the vacuum is invariant under all the BMS generators. Such a state can only exist when all the central charges in the algebra vanish, which is indeed the case as we will show later.

Similarly, the induced R vacuum can be defined as
\begin{align}
{\defa}_n|0_I\rangle_R=0,\ \  \forall n\in \mathbb{Z}
\end{align}
which is also BMS invariant, satisfying \eqref{trivial}. As ${\defa}_0$ annihilates the vacuum,  $\gamma_0$ should be regarded as a creation operator, and thus the induced R vacuum is not degenerate.
As a consequence, in the induced representation the R sector differs from the NS sector only in the mode expansion.
These features are in contrast with the highest weight R vacua, which are degenerate and not invariant under the global BMS algebra.

The normal ordering can be defined as follows,
\begin{align}
:{\defa}_{n}{\defb}_{m}:&=-{\defb}_{m}{\defa}_{n}\\
:{\defb}_{n}{\defa}_{m}:&={\defb}_{n}{\defa}_{m}.
\end{align}
where $n$ is an integer in the R sector and half integer in the NS sector.
Using the above normal ordering prescription we learn that the central charges
\eq{\label{icc34}c^{I}_L=c^{I}_M=0} for BMS algebra in the induced representation.

Define the propagators on the cylinder as
\begin{align}\label{oodef}
\langle O_{1}(\tau_1,\sigma_1)O_{2}(\tau_2,\sigma_2)\rangle \coloneqq T( O_{1}(\tau_1,\sigma_1)O_{2}(\tau_2,\sigma_2))-: O_{1}(\tau_1,\sigma_1)O_{2}(\tau_2,\sigma_2):
\end{align}
where the \(T\)-ordering is the time ordering on the cylinder
\begin{align} \label{RadOrd}
T\left(O_{1}(\tau_1,\sigma_1)O_{2}(\tau_2,\sigma_2)\right) &= \left\{\begin{array}{lll}
+O_{1}(\tau_1,\sigma_1)O_{2}(\tau_2,\sigma_2) & \text { for } & \tau_{1}>\tau_{2} \\
-O_{2}(\tau_2,\sigma_2)O_{1}(\tau_1,\sigma_1) & \text { for } & \tau_{1}<\tau_{2} \\
\end{array}\right.
\end{align}
The additional minus sign in the second line is due to the fermionic nature of the operators.
Under the definition \eqref{oodef}, the propagators for the fundamental fields $(\psi_1,\psi_2)$ can be calculated from the mode sum
\begin{align}
\langle \psi_1(\tau_1,\sigma_1)\psi_1(\tau_2,\sigma_2)\rangle&=0\nonumber
\\
\langle \psi_1(\tau_1,\sigma_1)\psi_2(\tau_2,\sigma_2)\rangle&=2\pi i\delta(\sigma_1-\sigma_2)(\theta(\tau_1-\tau_2)-\theta(\tau_2-\tau_1))\label{pp}\\
\langle \psi_2(\tau_1,\sigma_1)\psi_2(\tau_2,\sigma_2)\rangle&=4\pi i\delta'(\sigma_1-\sigma_2)(\theta(\tau_1-\tau_2)\tau_1+\theta(\tau_2-\tau_1)\tau_2)\nonumber
\end{align}
Note that these results are valid for both the NS and R sector. The reason is that there is no degenerate R vacua in this case and the discrete Fourier transformation for the delta-function admits both integer modes and half-integer modes as the basis.

As a consistency check, now we show that the propagators \eqref{pp} can also be derived from the path integral.
Using the fact that  the variation of the one-point functions with respect to the fields should vanish, namely
\begin{equation}
\frac{\delta}{\delta \psi_{1,2}}\langle \psi_{1,2}\rangle=0
\end{equation}
we get the following differential equations  for the propagator  \begin{equation}
\partial_{\sigma_1}\langle \psi_1(\tau_1,\sigma_1)\psi_1(\tau_2,\sigma_2)\rangle=0
\end{equation}
\begin{equation}
\partial_{\tau_1}\langle \psi_1(\tau_1,\sigma_1)\psi_1(\tau_2,\sigma_2)\rangle=0
\end{equation}
\begin{equation}
\partial_{\tau_1}\langle\psi_1(\tau_1,\sigma_1)\psi_2(\tau_2,\sigma_2)\rangle=2\pi i\delta(\sigma_1-\sigma_2)
\end{equation}
\begin{equation}
2\partial_{\sigma_1}\langle \psi_1(\tau_1,\sigma_1)\psi_2(\tau_2,\sigma_2)\rangle=\partial_{\tau_1}\langle \psi_2(\tau_1,\sigma_1)\psi_2(\tau_2,\sigma_2)\rangle
\end{equation}
It is then straight forward to check that the propagators in the induced vacuum \eqref{pp} are indeed solutions to the above equations for the Green's functions.
Note that the solutions \eqref{pp} feature delta-functions in the spatial directions, and theta-functions in the time directions,
properties also shared by the BMS scalar in the induced vacuum \cite{Hao:2021urq}, and predicted by general analysis \cite{Oblak:2015sea}.

For completeness, we also compute the correlation functions for the composite operator $P=\psi_1\psi_2$.
The two point function reads
\begin{equation}
\langle PP\rangle=-4\pi^2 \delta(\sigma_1-\sigma_2)
\end{equation}
The non-vanishing three point functions read
\begin{equation}
\begin{split}
\left\langle \psi_{1}\psi_{2}P\right\rangle =&-4\pi^{2}\delta(x_{1}-x_{3})\delta(x_{2}-x_{3})f_{13}f_{23}\\
\left\langle \psi_{2}\psi_{2}P\right\rangle =&-8\pi^{2}\delta(x_{2}-x_{3})\delta'(x_{1}-x_{3})f_{23}g_{13}\\
+&8\pi^{2}\delta(x_{1}-x_{3})\delta'(x_{2}-x_{3})f_{13}g_{23}
\end{split}
\end{equation}
where
\begin{equation}
f_{ij}=\theta(\tau_i-\tau_j)-\theta(\tau_j-\tau_i),\ \ g_{ij}=\tau_i\theta(\tau_i-\tau_j)+\tau_j(\theta(\tau_j-\tau_i))
\end{equation}
and $i,j=1,2,3$ label positions from left to right.

\subsection{State space}
Since all the ${\defa}_n$ modes are annihilation operators, the state space of the NS sector is spanned by
\begin{equation}\label{NSi}
|\vec{i}\rangle_{NS} :=\cdots {\defb}_{-1/2}^{i_{-1}}{\defb}_{1/2}^{i_{1}}\cdots|0_I\rangle_{NS}, \quad i_{n}=0,1.
\end{equation}
Similarly, the state space of the R sector is spanned by
\begin{equation}\label{Ri}
|\vec{i}\rangle_{R} :=\cdots {\defb}_{-1}^{i_{-1}}{\defb}_{0}^{i_{0}}{\defb}_{1}^{i_{1}}\cdots|0_I\rangle_{R}, \quad i_{n}=0,1.
\end{equation}
As there is no essential difference between the NS and R sectors, henceforth we use $|\vec{i}\rangle $ to denote both \eqref{NSi} and \eqref{Ri} and treat the two sectors in the same manner unless otherwise specified.
By acting the Virasoro zero mode $L_0$ on the states, we learn that the conformal weight is determined by the sum of the mode labels,
\begin{equation}
\Delta_{|\vec{i}\rangle}=-\sum_{i_k=1} k.
\end{equation}
Let us further define the ``length" $|i|$ of a state by the total number of $\gamma_k$ acting on the vacuum,
\eq{
|i|\equiv \sum_{k}i_k
}
From the commutation rules \begin{align}
 [L_n,{\defb}_m]=-\left(\frac{n}{2}+m\right){\defb}_{n+m},\quad
 [M_n,{\defb}_m]=-(n+2m){\defa}_{n+m},\nonumber
\end{align}
we learn that the action of  $M_n$ replaces one of the creation operator $\gamma_k$ by the annihilation operator $\beta_{n+k}$, which subsequently will either annihilate the vacuum and render a null state, or eliminate the operator $\gamma_{-(n+k)}$ if the latter exists.
As a result, $M_n$ either annihilate the state, or decrease the length by multiples of $2$.
This indicates that a generic state $|\vec i\rangle$ is not necessarily annihilated by $M_n$ for $n\neq0$, and thus is not a primary state as defined in \eqref{inducedm}.
In fact, any state annihilated by $M_n,\,\forall n\neq0$ can be obtained by a linear combination of the following states,
\eq{ {\defb}_{k_1}{\defb}_{-k_1}{\defb}_{k_2}{\defb}_{-k_2}\cdots \cdots{\defb}_{n}^{i_n}|0\rangle,\quad i_n=0,1\label{mn0}}
each of which is an eigenstate of $L_0$ with conformal weight $\Delta= -n$.
To further organize the states, we consider the action $M_0$. Note that $M_0$ acting on a state \eqref{mn0} will bring down one or more pairs of $\gamma_{-k}\gamma_k$, until finally reaching $|0\rangle$ or $\gamma_n|0\rangle$, and finally
\eq{
M_0|0\rangle&=M_0\gamma_n|0\rangle=0.
}
 This means that the states \eqref{mn0} can be organized into primary multiplets each satisfying \eqref{inducedm},
\eq{\label{primary1}
{\bf 1}&\sim\Big(|0\rangle, \gamma_{-k}\gamma_{k}|0\rangle,\gamma_{-k_1}\gamma_{k_1}\gamma_{-k_2}\gamma_{k_2}|0\rangle,\cdots \Big), \\
{\bf O_n}&\sim\Big(\gamma_n|0\rangle, \gamma_{-k}\gamma_{k} \,\gamma_n|0\rangle,\gamma_{-k_1}\gamma_{k_1}\gamma_{-k_2}\gamma_{k_2}\,\gamma_n|0\rangle,\cdots \Big), \label{{primaryn}}
}
where the identity multiplet ${\bf 1}$  includes the vacuum itself and other states created by multiple pairs of $\gamma_{-k}\gamma_{k}$, and the multiplet ${\bf O_n}$ includes $\gamma_n$ with all other creation operators paired up.
Therefore, we have infinite but countable numbers of primary multiplets and each has infinite rank.

The action of $L_n$ on the states \eqref{NSi} and \eqref{Ri} does not change the length $|i|$, but will shift the conformal weight by
$-n$,
\begin{equation}
\Delta(L_n|\vec{i}\rangle)=\Delta(|\vec{i}\rangle)-n
\end{equation}
In particular, acting $L_n,\, n\neq0$ on the primary multiplets will generate descendant states.

Let $H_{I}$ denote the module formed by the identity multiplet ${I}$ and its Virasoro descendants, and $H_{{\bf O}_n}$ denote the module generated from the primary multiplet ${\bf O}_n$.
Then the state space can be decomposed into different induced modules,
\begin{equation}
H=H_{I}\oplus\sum_{n}H_{{\bf O}_n}
\end{equation}
The states can be schematically described by Fig. \ref{inducedig}, where the round dots represent states, the black arrows represent the action of $L_n$, and the blue ones represent the action of $M_n$. A row represents states with the same conformal weight which increases from top to bottom. A column represents the states with the same length $|i|$, the number of creation operators $\gamma_n$. From left to right, the length of adjacent columns increases by $2$.
\begin{figure}[H]\centering
\label{inducedig}\begin{tikzcd}[scale cd = 1.2, column sep=small]
&{\color{black}\cdots}\arrow[dd, rightharpoonup, bend left=3, black]&&{\color{black}\cdots}\arrow[lldd, blue]\arrow[dd, rightharpoonup, bend left=3, black]&&&&{\color{black}\cdots}\arrow[dd, rightharpoonup, bend left=3, black]\arrow[lllldd, blue]\\
&&&&&&&\\
&{\color{black}\bullet}\arrow[dd, rightharpoonup, bend left=3, black]\arrow[uu, rightharpoonup, bend left=3, black]&&{\color{black}\bullet}\arrow[dd, rightharpoonup, bend left=3, black]\arrow[uu, rightharpoonup, bend left=3, black]\arrow[ll, blue]\arrow[lluu, blue]\arrow[lldd, blue]&&&&{\color{black}\bullet}\arrow[dd, rightharpoonup, bend left=3, black]\arrow[llll, blue]\arrow[lllluu, blue]\arrow[lllldd, blue]\arrow[uu, rightharpoonup, bend left=3, black]&&\cdots\\
&&&&&&&\\
&{\color{black}\cdots}\arrow[uu, rightharpoonup, bend left=3, black]&&{\color{black}\cdots}\arrow[lluu, blue]\arrow[uu, rightharpoonup, bend left=3, black]&&&&{\color{black}\cdots}\arrow[uu, rightharpoonup, bend left=3, black]\arrow[lllluu, blue]\\
\end{tikzcd}
\caption{Induced Module}\label{inducedig}
\end{figure}
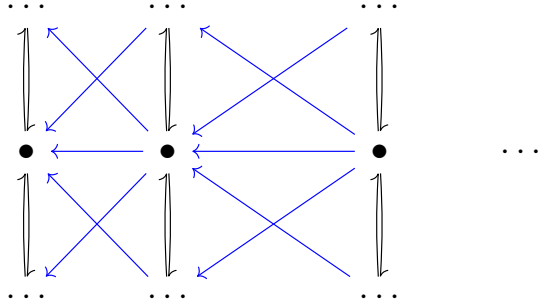

\subsection{Torus partition function}
The torus partition function for the induced vacuum can be calculated parallel to section 5. In particular, the trace should be taken over states \eqref{NSi} in the NS sector and \eqref{Ri} in the R sector, and we should also consider periodic and anti-periodic boundary conditions along the thermal circle.
The operator in $M_0$ does not contribute to the partition function,  and the computation reduces to the problem of finding the spectrum of $L_0$. The difference between the highest weight  and induced representations is the range of the eigenvalues of $L_0$, which is bounded from below in the former case, and  unbounded in the later one. In the induced representation,  any state $\gamma_{-k_1}\gamma_{-k_2}\cdots|0\rangle$ is paired with a partner $\gamma_{k_1}\gamma_{k_2}\cdots|0\rangle$, and the two states have opposite eigenvalues of $L_0$. Besides, the BMS fermions in the induced vacuum have vanishing central charges \eqref{icc34}, which do not contribute to the torus partition function.
Consequently, the torus partition function in the induced vacuum can be factorized and is given by,
\begin{equation}
Z_{R,R}(a,b)=Tr_R(-1)^F e^{-2\pi aL_{0}}=0
\end{equation}
\begin{equation}
Z_{R,NS}(a,b)=Tr_R e^{-2\pi aL_{0}}=\sqrt{\frac{\vartheta_{2}(ia)}{\eta(ia)}\frac{\vartheta_{2}(-ia)}{\eta(-ia)}}
\end{equation}
\begin{equation}
Z_{NS,NS}(a,b)=Tr_{NS}e^{-2\pi aL_{0}}=\sqrt{\frac{\vartheta_{3}(ia)}{\eta(ia)}\frac{\vartheta_{3}(-ia)}{\eta(-ia)}}
\end{equation}
\begin{equation}
Z_{NS,R}(a,b)=Tr_{NS}(-1)^F e^{-2\pi aL_{0}}=\sqrt{\frac{\vartheta_{4}(ia)}{\eta(ia)}\frac{\vartheta_{4}(-ia)}{\eta(-ia)}}
\end{equation}
where the factor like $\sqrt{\frac{\vartheta_{2}(ia)}{\eta(ia)}}$ comes from the contribution of states created by products of $\gamma_n$s with $n<0$, and the factor like $\sqrt{\frac{\vartheta_{2}(-ia)}{\eta(-ia)}}$ from those with $n>0$.
Note that the elliptic functions above are only formal, as they are only well defined on the upper half plane. The partition function of the induced sector is actually divergent, due to  the  unbounded spectrum of $L_0$.
These properties also appear in the BMS scalar model \cite{Hao:2021urq}.

\section{Supersymmetry}
In this section we discuss BMS invariant theories with supersymmetry by considering the free BMS scalar model \cite{Hao:2021urq} and free fermion models.
As discussed in section 2, both the BMS fermions \eqref{bmsfermion} and the chiral fermion \eqref{chiralfermion} are compatible with Carrollian symmetry, and hence can be used to construct symmetric models. We will discuss the two choices separately in the following.

In the following, we will use the Greek letters $\alpha,\beta,\lambda,\delta=1,2$ as spinor indices, which is raised and lowered by the charge conjugation matrix.  Whenever it does not cause confusion, we will omit the spinor indices when they are contracted. In that case, the northwest-southeast summation rule is assumed, for instance, $\bar{\chi}\sigma\equiv \bar{\chi}^{\alpha}\sigma_{\alpha}$.

\subsection{SUSY with BMS fermions}\label{susy bms ferm}
We first consider the free supersymmetric model built from the BMS fermion \eqref{bmsfermion}, together with the BMS scalar model
\begin{equation}\label{scalar}
S_{scalar} = \frac{1}{8\pi} \int e^{c}\wedge e^{d}\epsilon_{cd}g^{ab}\partial_{a}\phi\partial_{b}\phi =\frac{1}{4\pi} \int dx dy (\partial_{y}\phi)^{2}.
\end{equation}
The fermion field in \eqref{bmsfermion} has two real degrees of freedom, while the scalar field has only one degree of freedom. To match the number of degrees of freedom, we need to introduce an auxiliary scalar field $F$ which vanishes on-shell. The full supersymmetric action is thus given by
\begin{equation}
S=\frac{1}{8\pi} \int e^{c}\wedge e^{d}\epsilon_{cd}\left(g^{ab}\partial_{a}\phi\partial_{b}\phi+\bar{\psi}\Gamma^{a}\partial_{a}\psi-F^{2}\right).\label{susy}
\end{equation}
where the gamma matrices are given by \eqref{choiceI}. We will show that this model has $\mathcal N=2$ supersymmetry.
Written covariantly, the supersymmetric BMS theory takes a similar form as the free $\mathcal N=2$ supersymmetric QFT in 2 dimensions \cite{Green:2012oqa}.

The stress tensor in the bosonic part \cite{Hao:2021urq} and fermionic part are given by
\begin{equation}
\begin{split}
T_b&=-:\partial_x\phi\partial_y\phi:,\ \  M_b=-\frac{1}{2}:\partial_y\phi\partial_y\phi:,\\
T_f &= -\frac{1}{2}:\psi_{1}\p_{x}\psi_{2}:-\frac{1}{2}:\psi_{2}\p_{x}\psi_{1}:,\ \ M_f =-\frac{1}{2}:\psi_{1}\p_{y}\psi_{2}:.
\end{split}
\end{equation}
Then the total stress tensor read
\begin{equation}
T=T_b+T_f,\ \ M=M_b+M_f.
\end{equation}
The supersymmetry transformations with an infinitesimal spinor parameter $\epsilon_{\alpha}$ can be written in the covariant form
\begin{equation}
\begin{split}
\delta\phi&=i\epsilon^{\alpha}\psi_{\alpha}\\
\delta\psi_{\alpha}&=i(\Gamma^{a})_{\alpha}^{\ \ \beta}\epsilon_{\beta}\partial_{a}\phi-i\epsilon_{\alpha}F\\
\delta\bar{\psi}^{\alpha}&=-i\epsilon^{\beta}(\Gamma^{a})_{\beta}^{\ \ \alpha}\partial_{a}\phi-i\epsilon^{\alpha}F\\
\delta F&=-i\epsilon^{\alpha}(\Gamma^{a})_{\alpha}^{\ \ \beta}\partial_{a}\psi_{\beta}\label{susytran}
\end{split}
\end{equation}
As suggested by the similarity between the BMS SUSY action \eqref{susy} and its Lorentzian cousin, the above supersymmetric transformation also takes a similar form, except that the gamma matrices here are different.
By considering the variation of the action \eqref{susy} under the supersymmetry transformations \eqref{susytran}, we obtain a conserved supercurrent
\eq{
j^{a}_{\alpha}&=-\frac{i}{2\pi}(\Gamma^{b})^{\ \ \beta}_{\alpha}(\Gamma^{a})^{\ \ \lambda}_{\beta}\left(\partial_{b}\phi\right)\psi_{\lambda},\label{susycurr0}
}
where the superscript $a=1,2$ is a tangent space index, and the subscript $\alpha=1,2$ is a spinor space index.
The conserved current can be explicitly written in terms of components as,
\begin{align}
 j^{x}_1&=0,\quad j^x_2=-{1\over\pi}H\\
 j^{y}_1&={1\over 2\pi } H,\quad j^y_2={1\over 2\pi } \tilde G
 \end{align}
where \begin{equation}
H\equiv -i\partial_y\phi\psi_1,\ \ \ \tilde{G}\equiv -2i\partial_x\phi\psi_1-i\partial_y\phi\psi_2
\end{equation}
The conservation law $\partial_{a} j_\alpha^{a}=0$ can be expressed in components which reads
\begin{equation}
\begin{split}
&\partial_{y}H=0,\\
& \partial_{y}\tilde{G}-2\partial_{x}H=0,\label{conslaw}
\end{split}
\end{equation}
The first equation implies that $H$ is independent of $y$, while the second equation can be reorganized to define another $y-$independent field $G$,
\eq{
\p_y G=0,\quad G\equiv \tilde{G}-2y\partial_{x}H. \label{cl2}}
From the conservation laws \eqref{conslaw} \eqref{cl2}, we can construct two sets of conserved charges
\begin{equation}\label{supercharge}
Q_1[{\epsilon}]=\frac{1}{2\pi i}\oint \epsilon(x)H dx,\ \ Q_2[\epsilon]=\frac{1}{2\pi i}\oint\epsilon(x)G dx
\end{equation}
Among them, the zero modes are the global supercharges which generate the supersymmetry transformations,
\begin{equation}
Q_1\equiv Q_{1}[1]=\frac{1}{2\pi i}\oint H dx,\ \ Q_2\equiv Q_{2}[1]=\frac{1}{2\pi i}\oint G dx
\end{equation}
The supercharges generate transformations of free fields via the Poisson brackets,
\begin{equation}
\begin{split}
\delta \phi=&\{\epsilon^{\alpha}Q_{\alpha},\phi\}_{PB}=i\epsilon^{1}\psi_{1}+i\epsilon^{2}\psi_{2}\\
\delta \psi_{1}=&\{\epsilon^{\alpha}Q_{\alpha},\psi_{1}\}_{PB}=i\epsilon_{1}\partial_{y}\phi\\
\delta \psi_{2}=&\{\epsilon^{\alpha}Q_{\alpha},\psi_{2}\}_{PB}=-i\epsilon_{2}\partial_{y}\phi+2i\epsilon_{1}\partial_{x}\phi
\end{split}
\end{equation}
which agrees with the supersymmetry transformations \eqref{susytran} under the on-shell condition $F$=0.
Furthermore, the Poisson brackets between the two supercharges are given by
\begin{equation}
\begin{split}
i\{Q_{1},Q_{1}\}_{PB}=&0,\\
i\{Q_{1},Q_{2}\}_{PB}=&\frac{2}{2\pi i}\oint dxM=2P_{y},\\
i\{Q_{2},Q_{2}\}_{PB}=&\frac{4}{2\pi i}\oint dx\left(T-y\partial_{x}M\right)=4P_{x},\label{posbra1}
\end{split}
\end{equation}
where $P_{x},P_{y}$ are the conserved charges generating translations along the $x$ and $y$ direction respectively.

To study  more general charges in \eqref{supercharge}, we can perform mode expansion by taking the test function
\begin{equation}
\epsilon_r=x^{r+\frac{1}{2}},\quad r\in\mathbb{Z}+\frac{1}{2}
\end{equation}
so that we can define infinitely many charges
\begin{equation}\label{HGr}
H_r\equiv Q_{1}[\epsilon_r]=\frac{1}{2\pi i}\oint x^{r+\frac{1}{2}} H(x) dx,\ \ G_r\equiv Q_{2}[\epsilon_r]=\frac{1}{2\pi i}\oint x^{r+\frac{1}{2}} G(x) dx
\end{equation}
The mode expansion in the free BMS scalar is given by
\begin{equation}
\phi(x,y)=\sum_{n\in\mathbb{Z}} (A_n x^{-n}+y B_n x^{-n-1}),\label{scalarmode}
\end{equation}
 Plugging the mode expansions \eqref{scalarmode} and \eqref{ABmode} into \eqref{HGr}, we find
\begin{equation}
\begin{split}
H_{r}=&-i\sum_{m\in\mathbb{Z}}:\beta_{r-m}B_{m}:\\
G_{r}=&2i\sum_{m\in\mathbb{Z}}m:\beta_{r-m}A_{m}:-i\sum_{m\in\mathbb{Z}}:\gamma_{r-m}B_{m}:,\ \ \ r\in\mathbb{Z}+\frac{1}{2}
\end{split}
\end{equation}
We can then calculate the commutation relations between the supercharges and the modes,
\begin{equation}
[H_r,A_m]=i\beta_{r+m},\ \ [H_r,B_m]=0,\ \ \{H_r,\beta_s\}=0,\ \ \{H_r,\gamma_s\}=-iB_{r+s}
\end{equation}
\begin{equation}
[G_r,A_m]=i\gamma_{r+m},\ \ [G_r,B_m]=-2im\beta_{r+m},\ \ \{G_r,\beta_s\}=-iB_{r+s},\ \ \{G_r,\gamma_s\}=2i(r+s)A_{r+s}
\end{equation}
using which we get the following complete and closed supercharge algebra
\begin{equation}
[L_n,L_m]=(n-m)L_{n+m}+\frac{c_L}{12}n(n^2-1)\delta_{n+m,0}\nonumber
\end{equation}
\begin{equation}
[L_n,M_m]=(n-m)M_{n+m},\ \ \ [M_n,M_m]=0\nonumber
\end{equation}
\begin{equation}
[L_n,G_r]=\left(\frac{n}{2}-r\right)G_{n+r},\ \ [L_n,H_r]=\left(\frac{n}{2}-r\right)H_{n+r},\ \ [M_n,G_r]=2\left(\frac{n}{2}-r\right)H_{n+r}\label{susyI}
\end{equation}
\begin{equation}
[M_n,H_r]=0,\ \ \{H_r,H_s\}=0,\ \ \{G_r,H_s\}=2M_{r+s}\nonumber
\end{equation}
\begin{equation}
\{G_r,G_s\}=4L_{r+s}+\frac{2c_L}{3}(r^2-\frac{1}{4})\delta_{r+s,0}\nonumber
\end{equation}
where $c_L=3$ for the highest weight vacuum and the last three Lie brackets agree with the Poisson brackets in \eqref{posbra1}.
The algebra \eqref{susyI} is an $\mathcal{N}=2$ superalgebra containing two supercharges, with BMS algebra as the bosonic part.
The above algebra agrees with that in \cite{Mandal:2010gx} which was obtained from the UR limit of relativistic SCFTs, and that in \cite{Bagchi:2017cte} which was obtained classically from the supersymmetry generators in the superspace formalism, up to some coefficients related to the relative normalization between $\psi_1$ and $\psi_2$.

To conclude, we find that the action \eqref{susy} that includes a free BMS scalar \eqref{scalar} and two real fermions with the action \eqref{bmsfermion} has $\mathcal{N}=2$  supersymmetry.

\subsection{SUSY with chiral fermions}
In this subsection, we consider the chiral fermion case \eqref{chiralfermion}. The supersymmetric action can still be written in the covariant form \eqref{susy} but the gamma matrices and charge conjugation should be replaced by \eqref{chiralgamma}. The fermionic action now becomes
\begin{align}
S_f = \frac{1}{8\pi} \int e^{c}\wedge e^{d}\epsilon_{cd} \bar{\psi}\Gamma^{a}\partial_{a}\psi=\frac{1}{4\pi}\int dx dy (\psi_1\partial_{y}\psi_1+\psi_2\partial_y\psi_2).
\end{align}
In this case, there are two fundamental fermion fields $\psi_1$ and $\psi_2$, each of which behaves as a chiral fermion in a CFT$_2$. The fermions are invariant under the boost symmetry, and furthermore is annihilated by all the $M_n$ generators. The symmetry algebra is thus one copy of the Virasoro algebra only. The stress tensor for the fermion is given by
\begin{align}
T_f = -\frac{1}{2}:\psi_{1}\p_{x}\psi_{1}:-\frac{1}{2}:\psi_{2}\p_{x}\psi_{2}:,\ \ \ M_f =0,
\end{align}
with the following mode expansion
\begin{equation}
L^{(f)}_{n}=\sum_{m\in\mathbb{Z}+\frac{1}{2}}\frac{1}{2}\left(m+\frac{1}{2}\right)\left(:\beta_{n-m}\beta_{m}:+:\gamma_{n-m}\gamma_{m}:\right),\ \ \ M^{(f)}_{n}=0,
\end{equation}
where $\beta_{m}$ corresponds to the mode of $\psi_{1}$, and $\gamma_m$ for $\psi_2$,
with the anti-commutation relation $\{\beta_{m},\beta_{n}\}=\{\gamma_{m},\gamma_{n}\}=\delta_{m+n,0},\, \{\beta_{m},\gamma_{n}\}=0$.

As the supersymmetry transformations \eqref{susytran} and super currents \eqref{susycurr0} were calculated in the covariant way, they are still valid for the chiral fermions except that the spinor indices are raised or lowered by the identity matrix.
Plugging the gamma matrices \eqref{chiralgamma} into the super currents \eqref{susycurr0}, we find the four components of the super currents are
\eq{
j^{x}_{\alpha}&=0,\ \ \ j_{\alpha}^{y}=\frac{1}{2\pi}\left[\begin{array}{c}
H\\
G\end{array}\right].
}
where \eq{
H&=-i\partial_{y}\phi\psi_{1},\ \ G=-i\partial_{y}\phi\psi_{2}
}
The conservation law reads,
\begin{equation}
\partial_yH=0,\ \ \partial_y G=0
\end{equation}
Then we can get the global supercharges
\begin{equation}
\begin{split}
Q_{1}&=\frac{1}{2\pi i}\oint dx H,\\
Q_{2}&=\frac{1}{2\pi i}\oint dx G.
\end{split}
\end{equation}
One can also check the Poisson brackets between supercharges and free fields agree with the supersymmetry transformations \eqref{susytran} under on-shell condition
\begin{equation}
\begin{split}
\delta \phi=&\{\epsilon^{\alpha}Q_{\alpha},\phi\}_{PB}=i\epsilon^{1}\psi_{1}+i\epsilon^{2}\psi_{2}\\
\delta \psi_{1}=&\{\epsilon^{\alpha}Q_{\alpha},\psi_{1}\}_{PB}=i\epsilon_{1}\partial_{y}\phi\\
\delta \psi_{2}=&\{\epsilon^{\alpha}Q_{\alpha},\psi_{2}\}_{PB}=i\epsilon_{2}\partial_{y}\phi
\end{split}
\end{equation}
The Poisson brackets of the two supercharges read
\begin{equation}
i\{Q_{1},Q_{1}\}_{PB}=i\{Q_{2},Q_{2}\}_{PB}=2P_y,\ \ \ i\{Q_{1},Q_{2}\}_{PB}=0\label{posbra2}
\end{equation}
where again $P_y$ is the conserved charge generated by the translation on the $y$ direction as defined in \eqref{posbra1}.

We can also perform the mode expansion to find supercharge algebra,
\begin{equation}
\begin{split}
H_{r}&=-i\sum_{m\in\mathbb{Z}}:\beta_{r-m}B_{m}:,\\
G_{r}&=-i\sum_{m\in\mathbb{Z}}:\gamma_{r-m}B_{m}:.
\end{split}
\end{equation}
In this case, the commutation relations between supercharge modes and free field modes are
\begin{equation}
\begin{split}
[G_{r},A_{m}]=i\gamma_{r+m},&\ \ \ [G_{r},B_{m}]=0,\\
[H_{r},A_{m}]=i\beta_{r+m},&\ \ \ [H_{r},B_{m}]=0,\\
\{G_{r},\beta_{s}\}=0,&\ \ \ \{G_{r},\gamma_{s}\}=-iB_{r+s},\\
\{H_{r},\beta_{s}\}=-iB_{r+s},&\ \ \ \{H_{r},\gamma_{s}\}=0
\end{split}
\end{equation}
Then the complete and closed supercharge algebra can be written as follows
\begin{equation}
[L_n,L_m]=(n-m)L_{n+m}+\frac{c_L}{12}n(n^2-1)\delta_{n+m,0}\nonumber
\end{equation}
\begin{equation}
[L_n,M_m]=(n-m)M_{n+m},\ \ [M_n,M_m]=0\nonumber
\end{equation}
\begin{equation}
[L_n,G_r]=\left(\frac{n}{2}-r\right)G_{n+r},\ \ [L_n,H_r]=\left(\frac{n}{2}-r\right)H_{n+r}\label{susyII}
\end{equation}
\begin{equation}
[M_n,G_r]=0,\ \ [M_n,H_r]=0\nonumber
\end{equation}
\begin{equation}
\{H_r,H_s\}=2M_{r+s},\ \ \{G_r,H_s\}=0,\ \ \{G_r,G_s\}=2M_{r+s}\nonumber
\end{equation}
where $c_L=3$ for the highest weight vacuum and the last three Lie brackets agree with the Poisson brackets in \eqref{posbra2}.
The algebra \eqref{susyII} is an $\mathcal{N}=2$ superalgebra containing two supercharges, with BMS algebra as the bosonic part.
Thus, from the free scalar field and the two free chiral  fermion fields \eqref{chiralfermion} we have constructed a model with $\mathcal{N}=2$ supersymmetry.
Although both have $\mathcal{N}=2$, the structure of the superalgebra \eqref{susyII} differs from that of \eqref{susyI}, due to different choices of the fermionic part. In fact, each of the two supercharges in \eqref{susyII} forms a $\mathcal{N}=1$ subalgebra with the BMS generators,  acting on only one copy of the chiral fermions. This means that we can also construct an $\mathcal{N}=1$ symmetric theory by combining the BMS scalar \eqref{scalar} and one chiral fermion.

\subsection{Superspace method}
As an elegant method to construct supersymmetric models, superspace and superfield can also be applied to construct supersymmetric models with G/C symmetries. To do so, we use the two dimensional flat G/C spacetime \eqref{metric} as the base manifold,
so that a point in the superspace can be described by $(x^{a},\theta^{\alpha})$, with spacetime coordinates $x^a$ and Grassmann coordinates $\theta^{\alpha}$. The superfields are functions of the superspace coordinates which can be written as a power expansion in $\theta^{\alpha}$:
\begin{equation}
\Phi(x^{a},\theta^{\alpha})=\phi+i\theta^{\alpha}\psi_{\alpha}+\theta^{\alpha}\theta_{\alpha}F,\label{superfield}
\end{equation}
where the bosonic field $F$ with dimension 1 is an auxiliary field, $\phi$ is the scalar field and $\psi_{\alpha}$ is the two components spinor field. The supersymmetry algebra can be simply expressed as
\begin{equation}
\{Q_{\alpha},Q_{\beta}\}=2(\Gamma^{a})_{\alpha\beta}P_{a},\label{susyalg}
\end{equation}
where $P_{a}$ is the conserved charge under the infinitesimal translation.  The most general group element can be written in the superspace coordinates as
\begin{equation}
G(\varepsilon^{a},\zeta^{\alpha})=\exp\left(-\varepsilon^{a}P_{a}+\zeta^{\alpha}Q_{\alpha}\right)
\end{equation}
where $\varepsilon^{a}=(\varepsilon,\tilde{\varepsilon})$ parameterize the amount of spacetime translations, and $\zeta^{\alpha}$ parameterizes translation in the supercoordiate $\theta^\alpha$.
Now let us consider the effect of $Q_\alpha$, whose corresponding group element is given by $G(0,\eta^\alpha)$.
Multiplying $G(0,\eta^\alpha)$ to the left of another group element $G(\varepsilon^{a},\zeta^{\alpha})$ can be realized as a coordinate transformation in the superspace,
\begin{equation}
G(0,\eta^{\alpha})G(\varepsilon^{a},\zeta^{\alpha})=G(\varepsilon^{a}+\eta^{\alpha}(\Gamma^a)_{\alpha\beta}\zeta^{\beta},\zeta^{\alpha}+\eta^{\alpha})
\end{equation}
This effect of  coordinate transformation in the superspace can be generated by a differential operator $\hat{Q}_{\alpha}$
\begin{equation}\label{hQa}
\hat{Q}_{\alpha}=\frac{\partial}{\partial\theta^{\alpha}}+\theta^{\beta}(\Gamma^a)_{\beta\alpha}\partial_{a}.
\end{equation}
which indeed forms the supersymmetry algebra \eqref{susyalg} with $P_a=\p_a$ as the translational operator in spacetime.
The supersymmetry transformation of the superfield $\Phi$ can thus be obtained by acting on it with the supercharge \eqref{hQa},
\begin{equation}
\delta\Phi=\epsilon^{\alpha}\hat{Q}_{\alpha}\Phi=\epsilon^{\alpha}\left(\frac{\partial\Phi}{\partial\theta^{\alpha}}+\theta^{\beta}(\Gamma^a)_{\beta\alpha}\partial_{a}\Phi\right)=\delta\phi+i\theta^{\alpha}\delta\psi_{\alpha}+\theta^{\alpha}\theta_{\alpha}\delta F,
\end{equation}
whose components are indeed the supersymmetry transformation \eqref{susytran}.

To construct an action, we need to find covariant derivatives on the superspace by requiring that the covariant derivative transforms covariantly under coordinate transformation
\begin{equation}
\begin{split}
x'^{a}=&x^{a}+\eta^{\alpha}(\Gamma^{a})_{\alpha\beta}\theta^{\beta},\\
\theta'^{\alpha}=&\theta^{\alpha}+\eta^{\alpha}
\end{split}
\end{equation}
Note that the derivative $\partial/\partial \theta^{\alpha}$ does not transform covariantly
\begin{equation}
\frac{\partial}{\partial\theta^{\alpha}}=\frac{\partial\theta'^{\beta}}{\partial\theta^{\alpha}}\frac{\partial}{\partial\theta'^{\beta}}+\frac{\partial x'^{a}}{\partial\theta^{\alpha}}\frac{\partial}{\partial x'^{a}}=\frac{\partial}{\partial\theta'^{\alpha}}-\eta^{\beta}(\Gamma^{a})_{\alpha\beta}\frac{\partial}{\partial x'^{a}}
\end{equation}
Hence, the consistent covariant derivatives should be
\begin{equation}
\begin{split}
D_{\alpha}&=\frac{\partial}{\partial\theta^{\alpha}}-\theta^{\beta}(\Gamma^{a})_{\alpha\beta}\partial_{a}
\\\bar{D}^{\alpha}&=\frac{\partial}{\partial\theta_{\alpha}}-(\Gamma^{a})^{\alpha\beta}\theta_{\beta}\partial_{a}\label{covder}
\end{split}
\end{equation}
Now we can construct the free field supersymmetric action with the covariant derivatives \eqref{covder} and superfields \eqref{superfield} as follows
\begin{equation}
\begin{split}
S&=\frac{1}{32\pi}\int e^{a}\wedge e^{b}\epsilon_{ab} d\theta^{\alpha}d\theta^{\beta}\epsilon_{\alpha\beta}\left(\bar{D}^{\alpha}\bar{\Phi}D_{\alpha}\Phi\right)_{\theta^{\alpha}\theta_{\alpha}}\\
&=\frac{1}{4\pi}\int dxdy\left(g^{ab}\partial_{a}\phi\partial_{b}\phi+\bar{\psi}\Gamma^{a}\partial_{a}\psi-F^{2}\right),
\end{split}
\end{equation}
where the lower $\theta^{\alpha}\theta_{\alpha}$ means that we only take the terms proportional to $\theta^{\alpha}\theta_{\alpha}$. We see that the supersymmetric action turns out to be the one we have constructed in section \ref{susy bms ferm}.

\section*{Acknowledgments}

We would like to thank Bin Chen, Reiko Liu, Wenxin Lai, Jie-qiang Wu, Zhe-Fei Yu, Yuan Zhong for useful discussions. PH would like to thank the Institute of Theoretical Physics, Chinese Academy of Sciences (ITP) seminar and the ``2022 BIMSA Workshop on string theory ” at TSIMF, where some partial results of this work were presented and helpful comments were received. The work is partially supported by National Natural Science Foundation of China NO. 11735001, national key research and development program of China NO. 2020YFA0713000, and  Beijing Municipal Natural Science Foundation NO. Z180003.
\vspace{1cm}

\bibliographystyle{JHEP}
\bibliography{BMSFermion}

\end{document}